\documentclass[11pt,graphicx,amsmath]{article}
\usepackage{amsmath}
\usepackage{graphicx}
\usepackage{bm}
\usepackage{color}
\usepackage{amssymb}
\usepackage{amsfonts}
\usepackage{comment}
\usepackage{cite}
\usepackage{todonotes}
\usepackage{caption}
\usepackage{subcaption}

\def\bl#1\el{\begin{align}#1\end{align}}

\title{ The equation and solution of 4-point correlation function
         of galaxies in Gaussian approximation and its parity-odd part }

\author{\small Yang Zhang $^1$  \thanks{yzh@ustc.edu.cn~;~ ORCID: 0000-0001-7484-0043}  \,
               and  Shu-Guang   Wu $^{1, 2}$  \thanks{wusg@mail.ustc.edu.cn}   \\
\small {$^1$}  Department of  Astronomy,  Key Laboratory
	for Researches in Galaxies and Cosmology, \\
\small  University of Science and Technology of China, Hefei, Anhui, 230026, China \\
 \small {$^2$}   College of Modern Science and Technology,  \\
 \small China Jiliang University, Yiwu, Zhejiang, 322000, China  \\
}

\date{}

\def\be{\begin{equation}}
\def\ee{\end{equation}}
\def\ba{\begin{eqnarray}}
\def\ea{\end{eqnarray}}
\def\nn{\nonumber}
\def\bl#1\el{\begin{align}#1\end{align}}

\def\la{\langle}
\def\ra{\rangle}

\allowdisplaybreaks

\sf

\begin{document}
	
\maketitle

\large

\begin{abstract}

\large

Starting with the density field equation of a self-gravity fluid
 in a static Universe,
using the Schwinger functional differentiation technique,
we derive the field equation of the 4-point correlation function (4PCF)
of galaxies  in the Gaussian approximation,
which contains hierarchically 2PCF and  3PCF.
By use of the known solutions of 2PCF and 3PCF,
the equation of 4PCF becomes an inhomogeneous,  Helmholtz equation,
and contains only two physical parameters:
the mass $m$ of galaxy and the Jeans wavenumber $k_J$,
like the equations of the 2PCF and 3PCF.
We obtain the analytical solution of 4PCF
that consists of four portions, $\eta= \eta^0_{odd} + \eta^0_{even}  +\eta^{FP}  +\eta^I$,
and has a very rich structure.
$\eta^0_{odd}$ and $\eta^0_{even}$ form the homogeneous solution
and depend on boundary conditions.
The parity-odd $\eta^0_{odd}$ is more interesting
and qualitatively explains the observed parity-odd data of BOSS CMASS,
the parity-even $\eta^0_{even}$ contains
the disconnected 4PCF $\eta^{disc}$
(arising from a Gaussian random process),
and both $\eta^0_{odd}$ and $\eta^0_{even}$
are prominent at large scales  $r\gtrsim 10$Mpc,
and exhibit radial oscillations determined by the Jeans wavenumber.
$\eta^{FP}$ and $ \eta^I$ are parity-even,
and form the inhomogeneous solution.
$\eta^{FP}$ is the same as the Fry-Peebles ansatz for  4PCF,
and dominates at small scales  $r \lesssim  10$Mpc.
$\eta^I$ is an integration of the inhomogeneous term, subdominant.
We also compare the parity-even 4PCF with the observation data.

\end{abstract}


\section{Introduction }

The $n$-point correlation functions  (nPCF)
are important physical quantities in study of the distribution of galaxies
\cite{TotrujiKihara1969,PeeblesGroth1975,GrothPeebles1977,FryPeebles,Peebles1980,Peebles1993,
Fry1983,Fry1984a,Fry1984b,Fry1993},
and contain not only the statistical information of galaxies,
but also the physics of a system of self-gravity density field.
Pure statistical models will not be sufficient to understand
the underlying gravitational dynamics of the galaxy correlation.
The field equations of nPCF are fundamental
to analytically predict the physical properties of the system of galaxies.
Davis and Peebles  \cite{DavisPeebles1997}
treated the system of galaxies as a many-body system,
adopted the BBGKY (Bogoliubov-Born-Green-Kirkwood-Yvon) kinetic method,
worked with the Liouville's equation of probability function in the phase space,
and derived a set of five equations of 2PCF and velocity dispersions of galaxies.
But the initial conditions for the five unknowns are hard to specify in practice,
and moreover, the  equation of 2PCF was not closed.
Similar studies were made on the 3PCF
without a closed equation  \cite{Fry1984a,Fry1984b,Fry1993}.
In our previous studies,
we treated the system of galaxies by a self-gravity density field,
worked directly with the field equation of density fluctuation,
and employed the Schwinger external source method
\cite{Schwinger1951,Goldenfeld1992,BinneyDowrick1992,Zinn-Justin1996},
which has been commonly used in field theory to derive the equation of Green's functions.
We obtained the static equations of the 2PCF
in the Gaussian approximation \cite{Zhang2007},
the static nonlinear equation of 2PCF
\cite{ZhangMiao2009,ZhangChen2015,ZhangChenWu2019},
the evolutional nonlinear equation of 2PCF in the expanding Universe \cite{ZhangLi2021},
the static equation of 3PCF in the Gaussian approximation \cite{ZhangChenWu2019},
and the static nonlinear equation of 3PCF  \cite{WuZhang2022-8,WuZhang2022-2}.
These equations are closed and contain $m$ and $k_J$
as two physical parameters.
The equations in the Gaussian approximation have been solved analytically,
the nonlinear equations have been solved numerically,
and  the solutions  coherently explain
several seemingly-unrelated  prominent features of the observed 2PCF and 3PCF
 of the system of galaxies.

Fry and Peebles introduced the 4PCF of galaxies
and assumed  its form as a sum of products of 2PCF \cite{FryPeebles,Fry1983},
in analogy to the Groth-Peebles ansatz for 3PCF \cite{PeeblesGroth1975,GrothPeebles1977}.
Extensions to nPCF
\cite{Fry1983,Schaeffer1984,MeiksinSzapudiSzalay1992},
and generalizations to clusters \cite{HamiltonGott1988,SzapudiSzalayBoschan1992}
were made.
In lack of the closed equations of nPCF,
these works searched for possible phenomenological relations
within the hierarchical clustering picture.
An important property of the 4PCF is that it is parity sensitive.
The recent observation of BOSS CMASS \cite{Philcox2021,Philcox2022,Hou2022}
indicates that the 4PCF of galaxies contains the parity-odd signals,
which will be a cosmological evidence of parity violation
that was first found in particle physics \cite{LeeYang1956,WuCS1957}.

In this paper we shall derive the equation of the 4PCF
   in the Gaussian approximation, give the analytical solution,
   and compare
   with the observed 4PCF from BOSS CMASS \cite{Philcox2021,Philcox2022,Hou2022}.
In Sect 2 we give the derivation of the equation of 4PCF,
      and list also the equations and solutions of 2PCF and 3PCF in Gaussian approximation
      that will occur hierarchically in the equation of 4PCF.
Sect 3  we present the 4PCF solution and analyze the structure of its four portions.
In Sect 4 we use the parity-odd 4PCF solution to explain the parity-odd data from BOSS CMASS,
        and also compare the connected parity-even 4PCF solution
        with the corresponding data.
Sect 5 demonstrates the behaviors of three pieces
   of the parity-even solution in two simple configurations.
Sect 6 gives conclusion and discussions.

\section{The field equation of 4PCF in Gaussian approximation}

A self-gravity fluid is described by a set of equations,
including the continuity  equation, the Euler equation, and
the Poisson equation  \cite{Landau-Lifshitz,Peebles1980,ZhangLi2021}
\bl
\frac{\partial \rho}{\partial t} +\nabla \cdot (\rho {\bf v})
& =0,
\nn \\
\frac{\partial \bf v}{\partial t}+({\bf v}\cdot \nabla){\bf v}
& =-\frac{1}{\rho}\nabla p + \nabla \Phi,
\nn \\
\nabla^2 \Phi & =-4\pi G \rho \, .
\nn
\el
For the hydrostatical case, $\dot\rho=0$ and ${\bf v}=0$,
this set of equations leads to  the equation of the density field
\cite{Zhang2007,ZhangMiao2009,ZhangChen2015,ZhangChenWu2019}
\be \label{psifieldequ}
\nabla^2 \psi-\frac{(\nabla \psi)^2}{\psi}+k_J^2 \psi^2+J \psi^2 =0 ,
\ee
where $\psi(\mathbf{r}) \equiv  \rho(\mathbf{r}) / \rho_0$
is the rescaled dimensionless mass density
with $\rho_0$ being the mean mass density,
and  $k_J \equiv (4\pi G \rho_0/c_s^2)^{1/2}$ is the Jeans wavenumber,
$c_s $ is the sound speed, and the external source $J$ is introduced for
facilitating the  functional differentiations \cite{Schwinger1951,Goldenfeld1992}.
Eq.\eqref{psifieldequ} without $J$ can be also written as
 the known equation  \cite{Emden1907,Ebert1955,Antonov62,Bonnor1956,Lynden-Bell-Wood1968}
\be \label{Lane-Emden}
\nabla^2\phi + k^2_J e^{\phi}=0 ,
\ee
where $\phi \equiv  \ln \psi$ is the gravitational potential.
In  study of the correlation functions of galaxies,
it is convenient to work with the density field $\psi$,
instead of the potential $\phi$.
In cosmology, the density field $\psi$ in Eq.\eqref{psifieldequ}
is actually a statistical field,
which can be described by a grand partition function
\be  \label{Zdef}
Z [J]= \int D \psi \exp[-\alpha \int {\rm d}^3 {\bf r} \mathcal{H}(\psi, J)],
\ee
where $\alpha = \frac{c_s^2}{4\pi G m}$,
and the effective Hamiltonian is
$\mathcal{H}(\psi, J)
= \frac{1}{2} \big(\frac{\nabla \psi}{\psi}\big)^2 - k_J^2 \psi - J \psi$.
In field theory,  $Z [J]$ is also called
the generating functional for the correlation functions
of  the field $\psi$.
The connected nPCF  of $\delta \psi$  is defined by
\cite{Schwinger1951,Goldenfeld1992,BinneyDowrick1992,Zinn-Justin1996}
\bl \label{npcf}
G^{(n)}({\bf r}_1,\cdots,{\bf r}_n)
& =\la \delta \psi({\bf r}_1) \cdots \delta \psi({\bf r}_n)\ra
\nn
\\
& =\frac{1}{\alpha^n}\frac{\delta^n \log Z[J]}{\delta J({\bf r}_1)
	\cdots\delta J({\bf r}_n)} \bigg\vert_{J=0}
=\frac{1}{\alpha^{n-1}}\frac{\delta^{n-1} \la \psi({\bf r}_1) \ra }
{\delta J({\bf r}_2)\cdots\delta J({\bf r}_n)} \bigg\vert_{J=0}
          \, ,
\el
where $\delta \psi({\bf r}) = \psi({\bf r}) - \la \psi({\bf r}) \ra$
is the fluctuation of $\psi$ around the mean value $\la \psi({\bf r}) \ra$.
The collection of  $G^{(n)}$ for $n=2,3,4,...,$
provides systematic measures of the distribution of galaxies and clusters.
Since self-gravity enters  the field equation \eqref{psifieldequ}
and the partition function \eqref{Zdef},
$G^{(n)}$ is not only a statistical tool,
but also is influenced by the physics of the self-gravity density field.
So $G^{(n)}$  is a statistical and dynamical quantity.
Let us examine the generic, geometrical features of nPCF.
By definition \eqref{npcf},  $G^{(n)}$ has the symmetry under permutations
${\bf r}_i\leftrightarrow {\bf r}_j$ for $i,j=1,2,..., n$.
In applications, it is generally assumed that
$G^{(n)}$ is statistically homogeneous and  isotropic
 (invariant under simultaneous rotations of ${\bf r}_1, ... ,{\bf r}_n$).
These assumptions are also consistent with the isotropy and homogeneity of
the Universe on large scales.
From geometric perspective, the configuration of 2PCF
is a line formed  by the two points,
the parity operation $\mathbb{P} : {\bf r}_i\rightarrow -{\bf r}_i$ for $i=1,2$
on the line can be effectively replaced by a rotation.
The configuration of 3PCF is a triangle,
the parity operation on the triangle can be also replaced by a rotation
in 3-dim space.
The configuration of 4PCF is a tetrahedron,
the parity operation on the tetrahedron, however,
can not be replaced by rotations and translations  in 3-dim space.
So the 4PCF can record the parity \cite{Shiraishi2016}
and serve as a probe to the possible parity-odd information
encoded the distribution of galaxies in the Universe.
Recent surveys indicate that the observed 4PCF contains the parity-odd signals
\cite{Philcox2022,Hou2022,Philcox2021}.
As we shall demonstrate,
the solution of the equation of  $G^{(4)}$ contains a parity-odd part
that naturally  corresponds to the observed one.

To derive the field equations of  $G^{(n)}$,
we adopt the Schwinger external source method
and start with the ensemble average of Eq.(\ref{psifieldequ})
in the presence of an external source  $J$ \cite{Schwinger1951,Goldenfeld1992,Zhang2007}
\be \label{fieldequav2}
\Big\la \nabla^2 \psi-\frac{(\nabla \psi)^2}{\psi}
+k_J^2 \psi^2+J \psi^2 \Big\ra_{J}=0,
\ee
with $\la ... \ra$ denoting the ensemble average.
Taking functional derivative $(n-1)$ times of Eq.\eqref{fieldequav2}
with respect to the source $J$  and setting $J=0$
will lead to the equation of $G^{(n)}$.	
We shall derive respectively the equations of
$G^{(2)}$,  $G^{(3)}$, $G^{(4)}$ in the following.

First, consider $G^{(2)}$.
Taking  functional differentiation on \eqref{fieldequav2}
 with respect to $J$ once
\ba \label{2pcffielda}
\frac{1}{\alpha}\frac{\delta}{\delta J(\mathbf{r'})}
\bigg\la \nabla^2 \psi-\frac{(\nabla \psi)^2}{\psi}
+k_J^2 \psi^2+J \psi^2 \bigg\ra_J=0,
\ea
and setting $J=0$,
using the definition of the 2PCF
\bl \label{G2cal}
\frac{1}{\alpha } \frac{\delta}{\delta J(\mathbf{r'})  }
\la \psi({\bf r}) \ra \Big \vert_{J=0}
= G^{(2)}(\mathbf{r, r'} ),
\el
carrying out functional differentiation  on each term,
we obtain  the equation of the 2PCF (also called the Green's function in field theory)
in the Gaussian approximation \cite{Zhang2007}
\be  \label{Gaussian2PCF}
\nabla^2  G^{(2)}(\mathbf{r, r'})
	+ 2 k_J^2 G^{(2)}(\mathbf{r, r'})
=-\frac{1}{\alpha} \delta^{(3)}(\mathbf{r-r'}) ,
\ee
with  $\nabla^2= \nabla^2_{\bf r}$.
This is an inhomogeneous  Helmholtz equation with a delta source,
and contains $k_J$ and the mass $m$
as the two independent physical parameters of the fluid.
The  solution is (also denoted by $\xi$ in literature)
\bl \label{gsolution}
 G^{(2)}(\mathbf{r} , \mathbf{r}')=\xi(|{\bf r}-{\bf r'|})
& = A_\mathrm{m} \Big(
  d_0 \frac{\cos \big( k_0 \lvert \mathbf{r-r'} \rvert \big)}
{ \lvert \mathbf{r-r'} \rvert}
+ (1-d_0) \frac{\sin \big( k_0 \lvert \mathbf{r-r'} \rvert \big)}
{ \lvert \mathbf{r-r'} \rvert} \Big),
\el
where
\be \label{Adef}
A_\mathrm{m} = \frac{1}{4 \pi \alpha} = \frac{G m}{c_s^2}
\ee
is  the amplitude,   $k_0 = \sqrt{2} k_J$ is the reduced Jeans wavenumber,
$d_0$ is a constant to be determined by the boundary condition.
Fitting with the observed 2PCF indicates $d_0=1$,
$k_0 \simeq 0.05 h$ Mpc$^{-1}$ and $A_m \simeq 3\, h^{-1}\, $Mpc for galaxies
\cite{Zhang2007,ZhangMiao2009,ZhangChen2015,ZhangChenWu2019}.
As is seen, $m$ influences the clustering amplitude,
and $k_J$ determines the scale of the correlation.
The solution \eqref{gsolution} predicts that
more massive galaxies have a higher amplitude of correlation,
as long been observed and thought as a puzzling feature of clustering
\cite{HauserPeebles1973,BahcalSoneira1983,KlypinKopylov1983,
Bahcal1996,Bahcal2003}.
The phenomenological scaling of the correlation length
with the inter-galaxy separation \cite{BahcalWest1992,BahcalCen1992}
is actually implied by the solution \eqref{gsolution}.
More prominently, the solution \eqref{gsolution} also predicts that the 2PCF
is periodic oscillatory
with a wavelength  $2\pi /(\sqrt{2}k_J) \sim 100 h^{-1}$ Mpc.
As pointed out in Ref.\cite{ZhangLi2021},
this 100 Mpc feature is not an imprint of
the so-called sound horizon \cite{EisensteinHu1998,Eisenstein2005},
which is  in fact not an observable statistically
and has a value much larger than 100 Mpc.
These salient features have been confirmed by early observations
in galaxies \cite{Broadhurst1990,Broadhurst1995,Tucker1997}
and in clusters, \cite{Einasto1997,Einasto2002,Tago2002},
and by more recent observations with increasingly cumulative data
of galaxies \cite{Beutler6dFGS2011,SanchezSDSS2017,Anderson2014,RuggeriBlake2020,
KazinWiggleZ2014}
and of quasars \cite{Blake2011,Ata2018,Busca2013,BausticaBusca2017,Agathe2019}.
Higher order density fluctuations beyond the Gaussian approximation
will enhance $\xi$ at small scales
and yield a better description of observational data
\cite{ZhangMiao2009,ZhangChen2015,ZhangChenWu2019}.
The evolutionary equation of 2PCF in the expanding Universe
has been given by Ref.\cite{ZhangLi2021}.
The solution  \eqref{gsolution} is divergent as
$|{\bf r}-{\bf r}'|\rightarrow 0$,  like the Green functions of quantum fields
\cite{ZhangYeWang2020,ZhangWangYe2020,ZhangYePRD2022,YeZhangWang2022},
and we shall not discuss the issue here.

Beside the solution   \eqref{gsolution},
the equation \eqref{Gaussian2PCF}  is generally allowed
to have a  homogeneous solution
\bl
G^{(2)}_{0} ({\bf r, r'}) & = \sum_{l=1}^\infty  \sum_{m=-l}^{l}
 c_{lm}  Y_l^m ( \theta, \phi)
\Big[ d_l n_l (k_0 |{\bf r- r'}|) + c_l   j_l (k_0 |{\bf r- r'}|) \Big] ,
\label{G2homsol}
\el
which satisfies the homogeneous Helmhotz equation
$\left(\nabla^2_{\bf r}  +2 k_J^2\right) G^{(2)}_{0} (\mathbf{r}, {\bf r}') =0$,
where $ Y_l^m(\theta, \phi)$ are the spherical harmonic functions
for the direction of $\mathbf{r-r'}$,
and $j_l$  and $n_l$ are the spherical Bessel's functions.
The coefficients should satisfy $c_{lm}=0$ for $l=odd$,
as required by the symmetry under permutation.
Since the observed 2PCF from various surveys is  direction-independent,
the homogeneous solution \eqref{G2homsol} is set to zero $G^{(2)}_{0}=0$
as the boundary condition.

Next,   consider $G^{(3)}$.
Taking functional derivative of  \eqref{fieldequav2}
with respect to $J$  twice
\ba \label{4pcffielda}
\frac{1}{\alpha^2}\frac{\delta^2}{\delta J(\mathbf{r'})	\delta J(\mathbf{r''})}
\bigg\la \nabla^2 \psi-\frac{(\nabla \psi)^2}{\psi}
+k_J^2 \psi^2+J \psi^2 \bigg\ra_J=0,
\ea
and setting  $J=0$,
using the definition of the 3PCF
\ba\label{G3cal}
\frac{1}{\alpha^2} \frac{\delta^2}{\delta J({\bf r}_2) \delta J({\bf r}_3)}
\la \psi({\bf r}_1) \ra \Big  \arrowvert_{J=0}
=G^{(3)}({\bf r}_1, {\bf r}_2, {\bf r}_3),
\ea
performing  functional differentiation  on each term,
we obtain the field equation of the 3PCF in the Gaussian approximation
\cite{ZhangChenWu2019,WuZhang2022-2,WuZhang2022-8}
\bl\label{Gaussian3PCF}
\nabla^2  G^{(3)}({\bf r}_1, {\bf r}_2, {\bf r}_3)
	+ 2 k_J^2  G^{(3)}({\bf r}_1, {\bf r}_2, {\bf r}_3)
& =   2 \nabla G^{(2)}(r_{12}) \cdot \nabla G^{(2)}(r_{13})
- 2 k_J^2 G^{(2)}(r_{12}) G^{(2)}(r_{13})
\nn \\
& ~~ -\frac{2}{\alpha} \delta^{(3)}({\bf r}_1 -{\bf r}_2) G^{(2)}(r_{13})
	-\frac{2}{\alpha} \delta^{(3)}({\bf r}_1 -{\bf r}_3) G^{(2)}(r_{12}) ,
\el
where $\nabla^2 \equiv \nabla^2_{ {\bf r}_1 }$, $r_{ij}  \equiv |{\bf r}_i-{\bf r}_j|$,
and $G^{(2)}(r_{ij})$
satisfies \eqref{Gaussian2PCF} and is given by \eqref{gsolution}.
The equation \eqref{Gaussian3PCF}  is also a Helmholtz equation with
the inhomogeneous terms consisting of  $G^{(2)}$ and the delta source.
The 3PCF solution (also denoted by $\zeta$) is  \cite{ZhangChenWu2019}
\be  \label{GPansatz}
G^{(3)}({\bf r}_1, {\bf r}_2, {\bf r}_3)
=\zeta_{123}
= G^{(2)}(r_{12}) G^{(2)}(r_{23})
+G^{(2)}(r_{23}) G^{(2)}(r_{31})
+G^{(2)}(r_{31}) G^{(2)}(r_{12}) .
\ee
Amazingly,  this 3PCF solution in the Gaussian approximation
is just the same as
the Groth-Peebles ansatz (with $Q=1$) for 3PCF \cite{GrothPeebles1977,PeeblesGroth1975}.
As a function of three independent variables $r_{12}, r_{13}, r_{23}$,
the 3PCF solution \eqref{GPansatz} is symmetric under
permutations ${\bf r}_i\leftrightarrow {\bf r}_j$ with $i,j=1,2,3$,
invariant under the parity operation
$\mathbb{P} : {\bf r}_i \rightarrow {- \bf r}_i$,
as well as the spatial translations and rotations.
(If the equation \eqref{Gaussian3PCF} did not contain the delta source,
\eqref{GPansatz} would  not be the solution.)
Since the equation of 3PCF contains new information beyond that of 2PCF,
the boundary condition of 3PCF should be determined from
the observed 3PCF,
and the coefficient $d_0$ in \eqref{GPansatz} via $G^{(2)}$
generally differs  from what inferred from the observed 2PCF.

The equation \eqref{Gaussian3PCF} also has a general,  homogeneous solution
\be\label{G3homsol}
G^{(3)}_0 \left(\mathbf{r}_1, \mathbf{r}_2, \mathbf{r}_3 \right)
=y\left(\mathbf{r}_1\right) y\left(\mathbf{r}_2\right)y\left(\mathbf{r}_3\right)
 ,
\ee
where $y$ satisfies the equation
$\left(\nabla^2_{\bf r}  +2 k_J^2\right) y\left(\mathbf{r}\right) =0$
and is given by
\be\label{G3y}
y\left(\mathbf{r}\right)
= \sum_{l, m} C_{m l} Y_l^m(\hat {\bf r})
\left[ c_l j_l (k_0\left|\mathbf{r}\right| )
      + d_l n_l (k_0\left|\mathbf{r}\right| ) \right] .
\ee
The homogeneous solution \eqref{G3homsol}
has formally less symmetry than the  inhomogeneous solution \eqref{GPansatz}.
Various galaxy surveys so far indicate that the observed 3PCF
are independent of directions,
so we shall  set $G^{(3)}_{0}=0$ as the boundary condition in this paper.
If the observational data in future
are processed for a general direction-dependent 3PCF,
the form of the solution \eqref{G3homsol} can be employed for that purpose.
One might try another possible homogeneous solution of 3PCF
as a function of $({\bf r}_i -{\bf r}_j)$ as the following
\be
G^{(3)}_0 \left(\mathbf{r}_1, \mathbf{r}_2, \mathbf{r}_3 \right)
= y({\bf r}_1, {\bf r}_2) y({\bf r}_2, {\bf r}_3)
+ y({\bf r}_2, {\bf r}_3) y({\bf r}_3, {\bf r}_1)
+  y({\bf r}_3, {\bf r}_1) y({\bf r}_1, {\bf r}_2) ,
\nn
\ee
where $y({\bf r}_1, {\bf r}_2)$  is the same as  \eqref{G2homsol},
satisfying $ (\nabla^2_{\bf r_1}  +2 k_J^2 ) y({\bf r}_1, {\bf r}_2) =0$.
But, as can be checked, this is not a homogeneous solution of \eqref{Gaussian3PCF}.

As is known, the 3PCF of a Gaussian random  process is zero,
according to the Isserlis-Wick theorem \cite{Hristopulos,Isserlis1918,Wick1950}.
As such,  the  existence of a nonzero $G^{(3)}$ given by  \eqref{GPansatz}
indicates that the self-gravity
 density fluctuation described by the Gaussian approximation
is not a Gaussian random  process in statistics.
This is due to the presence of long-range gravity in the fluid.
The Gaussian approximation is the lowest order of approximation
to adequately account for the fluctuations of the self-gravity fluid.
(See more discussions in Sect 6.)
Beyond the Gaussian approximation,
the equation of $G^{(3)}$ will contain  more higher-order terms
than Eq.\eqref{Gaussian3PCF},
and the solution will be much more complicated than  \eqref{GPansatz}
(see Refs.\cite{WuZhang2022-8,WuZhang2022-2} for details).

Consider $G^{(4)}$.
Similarly, taking functional derivative with respect to $J$ thrice
\bl \label{4pcffielda}
\frac{1}{\alpha^3}\frac{\delta^3}{\delta J({\bf r}_2)
	\delta J({\bf r}_3) \delta J({\bf r}_4)}
\bigg\la \nabla^2 \psi-\frac{(\nabla \psi)^2}{\psi}
+k_J^2 \psi^2+J \psi^2 \bigg\ra_J=0,
\el
and setting $J=0$, using the definition of the 4PCF,
\be \label{def4pcf}
\frac{1}{\alpha^3}\frac{\delta^3}{\delta J({\bf r}_2)
	\delta J({\bf r}_3) \delta J({\bf r}_4)}
\la \psi({\bf r}_1) \ra \arrowvert_{J=0}=G^{(4)}({\bf r}_1,{\bf r}_2,{\bf r}_3,{\bf r}_4) ,
\ee
we obtain the field equation of the 4PCF  in Gaussian approximation
(see Appendix A for detailed derivation)
\bl \label{Gaussian4PCF}
& \nabla^2 G^{(4)}({\bf r}_1,{\bf r}_2,{\bf r}_3,{\bf r}_4)
+2 k_J^2 G^{(4)}({\bf r}_1,{\bf r}_2,{\bf r}_3,{\bf r}_4) \nonumber \\
& +2 G^{(2)}(r_{12}) \nabla G^{(2)}(r_{13})
  \cdot \nabla G^{(2)}(r_{14})
  + 2 G^{(2)}(r_{13})
   \nabla G^{(2)}(r_{14})
    \cdot \nabla G^{(2)}(r_{12})
\nn \\
& +2 G^{(2)}(r_{14})
\nabla G^{(2)}(r_{12})
\cdot \nabla G^{(2)}(r_{13})
 \nn \\
& -2 \nabla G^{(2)}(r_{12})
  \cdot \nabla G^{(3)}({\bf r}_1,{\bf r}_3,{\bf r}_4)
-2 \nabla G^{(2)}(r_{13})
    \cdot \nabla G^{(3)}({\bf r}_1,{\bf r}_2,{\bf r}_4)
    \nn \\
& -2 \nabla G^{(2)}(r_{14})
\cdot \nabla G^{(3)}({\bf r}_1,{\bf r}_2,{\bf r}_3)
\nn  \\
& +2 k_J^2 \bigg(G^{(3)}({\bf r}_1,{\bf r}_2,{\bf r}_4) G^{(2)}(r_{13})
+G^{(3)}({\bf r}_1,{\bf r}_3,{\bf r}_4) G^{(2)}(r_{12})
+G^{(3)}({\bf r}_1,{\bf r}_2,{\bf r}_3) G^{(2)}(r_{14}) \bigg)
  \nn  \\
 = & -\frac{2}{\alpha} \delta^{(3)}({\bf r}_1 -{\bf r}_2)
  \bigg( G^{(2)}(r_{13}) G^{(2)}(r_{14})
+ G^{(3)}({\bf r}_1,{\bf r}_3,{\bf r}_4) \bigg)
\nn \\
& - \frac{2}{\alpha} \delta^{(3)}({\bf r}_1 -{\bf r}_3)
  \bigg( G^{(2)}(r_{12}) G^{(2)}(r_{14})
+ G^{(3)}({\bf r}_1,{\bf r}_2,{\bf r}_4) \bigg)
  \nonumber \\
& - \frac{2}{\alpha} \delta^{(3)} ({\bf r}_1 -{\bf r}_4)
\bigg( G^{(2)}(r_{12}) G^{(2)}(r_{13})
+  G^{(3)}({\bf r}_1,{\bf r}_2,{\bf r}_3) \bigg) ,
\el
which contains $G^{(3)}$ and  $G^{(2)}$ hierarchically.
Using  $G^{(3)}$ in the Gaussian approximation given by \eqref{GPansatz},
the field equation  \eqref{Gaussian4PCF} becomes
\bl \label{solve4PCF-1}
&\nabla^2 G^{(4)}({\bf r}_1,{\bf r}_2,{\bf r}_3,{\bf r}_4)
+2 k_J^2  G^{(4)}({\bf r}_1,{\bf r}_2,{\bf r}_3,{\bf r}_4) \nonumber \\
=& 2 G^{(2)}(r_{12})
\nabla G^{(2)}(r_{13}) \cdot \nabla G^{(2)}(r_{14})
+2 G^{(2)}(r_{13})
\nabla G^{(2)}(r_{12}) \cdot \nabla G^{(2)}(r_{14})
\nn \\
& +2 G^{(2)}(r_{14})
\nabla G^{(2)}(r_{12}) \cdot \nabla G^{(2)}(r_{13})
\nn  \\
&+2 G^{(2)}(r_{34}) \bigg(
\nabla G^{(2)}(r_{12}) \cdot \nabla G^{(2)}(r_{13})
+\nabla G^{(2)}(r_{12}) \cdot \nabla G^{(2)}(r_{41})\bigg)
\nonumber \\
&+2 G^{(2)}(r_{24}) \bigg(
\nabla G^{(2)}(r_{13}) \cdot \nabla  G^{(2)}(r_{12})
+\nabla G^{(2)}(r_{13}) \cdot \nabla G^{(2)}(r_{41})\bigg)
\nonumber \\
&+2 G^{(2)}(r_{23}) \bigg(
\nabla G^{(2)}(r_{14}) \cdot \nabla  G^{(2)}(r_{12})
+\nabla G^{(2)}(r_{14}) \cdot \nabla G^{(2)}(r_{31})\bigg)
\nonumber \\
&-2 k_J^2 G^{(2)}(r_{13})
\bigg(G^{(2)}(r_{12}) G^{(2)}(r_{24})
+G^{(2)}(r_{24}) G^{(2)}(r_{41})
+G^{(2)}(r_{41}) G^{(2)}(r_{12}) \bigg)
\nonumber \\
&-2 k_J^2  G^{(2)}(r_{12}) \bigg( G^{(2)}(r_{13}) G^{(2)}(r_{34})
+G^{(2)}(r_{34}) G^{(2)}(r_{41})
+G^{(2)}(r_{41}) G^{(2)}(r_{13}) \bigg)
\nonumber \\
&-2 k_J^2  G^{(2)}(r_{14})
\bigg(G^{(2)}(r_{12}) G^{(2)}(r_{23})
+G^{(2)}(r_{23}) G^{(2)}(r_{31})
+G^{(2)}(r_{31}) G^{(2)}(r_{12}) \bigg)
 \nonumber \\
&-\frac{2}{\alpha} \delta^{(3)}({\bf r}-{\bf r}_2)
 \bigg( 2 G^{(2)}(r_{13}) G^{(2)}(r_{14})
+ G^{(2)}(r_{13}) G^{(2)}(r_{34})
+ G^{(2)}(r_{34}) G^{(2)}(r_{41}) \bigg)
 \nonumber\\
&- \frac{2}{\alpha} \delta^{(3)}({\bf r}-{\bf r}_3)
 \bigg( 2 G^{(2)}(r_{12}) G^{(2)}(r_{14})
+ G^{(2)}(r_{12}) G^{(2)}(r_{24})
+G^{(2)}(r_{24}) G^{(2)}(r_{41})  \bigg)
 \nonumber \\
&- \frac{2}{\alpha} \delta^{(3)} ({\bf r}-{\bf r}_4)
\bigg( 2 G^{(2)}(r_{12}) G^{(2)}(r_{13})
+ G^{(2)}(r_{12}) G^{(2)}(r_{23})
+G^{(2)}(r_{23}) G^{(2)}(r_{31}) \bigg) ,
\el
with the inhomogeneous terms being composed of
the known $G^{(2)}$  and the delta sources.
The structure of  equation \eqref{solve4PCF-1}
is of the Helmholtz type,
and resembles  the equations of 2PCF and 3PCF
\cite{Zhang2007,ZhangMiao2009,ZhangChen2015,ZhangChenWu2019,WuZhang2022-8,WuZhang2022-2}.
All these equations have two physical parameters:  $m$ and $k_J$.

\section{The  analytical solution of the 4PCF  in Gaussian approximation}

We shall give the analytical solution  $G^{(4)}$ of Eq.\eqref{solve4PCF-1}.
First we shall use  the following formulae
\be \label{identity02}
2 \nabla u \cdot \nabla v = \nabla^2(uv) -  v \nabla^2 u  -  u \nabla^2 v,
\ee
and
\be \label{identity01}
2 v \nabla u \cdot \nabla w + 2 u \nabla v \cdot \nabla w
+ 2 w \nabla u \cdot \nabla v
=\nabla^2(uvw) - v w \nabla^2 u  - u w \nabla^2 v - uv \nabla^2 w,
\ee
where $u$, $v$ and $w$ are arbitrary functions.
We apply the formula  \eqref{identity01} to
the first three terms $2 G^{(2)} \nabla G^{(2)} \cdot \nabla G^{(2)}$
on the r.h.s of \eqref{solve4PCF-1},
and apply the formula  \eqref{identity02} to other six terms $2 \nabla G \cdot \nabla G$
 on the rhs of  \eqref{solve4PCF-1},
and make use of the equation \eqref{Gaussian2PCF} of $G^{(2)}$.
Then, by regrouping and  simplifications,
Eq.\eqref{solve4PCF-1} is written in the following  simple form
\bl \label{g4g2}
& \big( \nabla^2 + 2 k_J^2 \big)
 \Big[ G^{(4)}({\bf r}_1,{\bf r}_2,{\bf r}_3,{\bf r}_4)
- \eta^{FP}({\bf r}_1,{\bf r}_2,{\bf r}_3,{\bf r}_4)  \Big]
= -2 k_J^2  G^{(2)}(r_{12})G^{(2)}(r_{13}) G^{(2)}(r_{14}),
\el
where
\bl \label{solh}
\eta^{FP} ({\bf r}_1,{\bf r}_2,{\bf r}_3,{\bf r}_4)
& \equiv  \Big[  G^{(2)}(r_{14}) G^{(2)}(r_{42}) G^{(2)}(r_{23})
+ G^{(2)}(r_{13}) G^{(2)}(r_{32}) G^{(2)}(r_{24})
  \nn \\
&  + G^{(2)}(r_{12}) G^{(2)}(r_{23}) G^{(2)}(r_{34})
 + G^{(2)}(r_{23}) G^{(2)}(r_{34}) G^{(2)}(r_{41})
 \nn \\
& + G^{(2)}(r_{34}) G^{(2)}(r_{42}) G^{(2)}(r_{21})
+ G^{(2)}(r_{24}) G^{(2)}(r_{43}) G^{(2)}(r_{31})
 \nonumber \\
& +   G^{(2)}(r_{23}) G^{(2)}(r_{31}) G^{(2)}(r_{14})
+   G^{(2)}(r_{24}) G^{(2)}(r_{41}) G^{(2)}(r_{13})
  \nn \\
&+ G^{(2)}(r_{21}) G^{(2)}(r_{14})G^{(2)}(r_{43})
+   G^{(2)}(r_{32}) G^{(2)}(r_{21}) G^{(2)}(r_{14})
  \nonumber \\
& +  G^{(2)}(r_{31}) G^{(2)}(r_{12}) G^{(2)}(r_{24})
 +   G^{(2)}(r_{21})G^{(2)}(r_{13}) G^{(2)}(r_{34})   \Big]
 \nn \\
& + \Big[   G^{(2)}(r_{12}) G^{(2)}(r_{13}) G^{(2)}(r_{14})
+  G^{(2)}(r_{23}) G^{(2)}(r_{24}) G^{(2)}(r_{21})
\nonumber \\
& +  G^{(2)}(r_{32}) G^{(2)}(r_{34}) G^{(2)}(r_{31})
+  G^{(2)}(r_{42}) G^{(2)}(r_{43}) G^{(2)}(r_{41}) \Big]
\el
consisting  of sixteen terms of products $G^{(2)} G^{(2)} G^{(2)}$
and each $G^{(2)}$ being given by \eqref{gsolution}.
Interestingly, the expression  \eqref{solh}  of $\eta^{FP}$
has the same form of the Fry-Peebles ansatz \cite{FryPeebles} for 4PCF
\bl \label{frypeeblesansatz}
&   R_a \big[ G^{(2)}(r_{12}) G^{(2)}(r_{23}) G^{(2)}(r_{34})
        +\mathrm{sym. (12 \, \, terms)} \big]
 \nn \\
& ~~  +R_b \big[ G^{(2)}(r_{12}) G^{(2)}(r_{13}) G^{(2)}(r_{14})
            +\mathrm{sym. (4 \, \, terms)} \big]
\el
with the constants $R_a=R_b=1$.
$\eta^{FP}$ in \eqref{solh} is symmetric under permutations
 ${\bf r}_i\leftrightarrow {\bf r}_j$ for $i,j=1,2,3,4$,
and invariant under the spatial translation and rotation, and is parity-even.
Denoting
\[
\eta^I ({\bf r}_1,{\bf r}_2,{\bf r}_3,{\bf r}_4)
  \equiv  G^{(4)}({\bf r}_1,{\bf r}_2,{\bf r}_3,{\bf r}_4)
        -\eta^{FP} ({\bf r}_1,{\bf r}_2,{\bf r}_3,{\bf r}_4)
      ,
\]
the equation \eqref{g4g2} is rewritten as
\be  \label{u-equ}
\big( \nabla^2 + 2 k_J^2 \big) \eta^I ({\bf r}_1,{\bf r}_2,{\bf r}_3,{\bf r}_4)
   = -2 k_J^2  G^{(2)}(r_{12}) G^{(2)}(r_{13}) G^{(2)}(r_{14}).
\ee
This is also  a Helmholtz  equation with an inhomogeneous term.
The solution of \eqref{u-equ} consists of
 a special solution $\eta^I$ and a general homogeneous solution $\eta^0$.
Thus,  the full 4PCF solution (denoted also by  $\eta$)
will be  a sum of three portions
\be \label{4pcfSol}
G^{(4)} = \eta =\eta^0 +\eta^I +\eta^{FP} .
\ee

\begin{figure}[htbp]
	\centering
	\includegraphics[width=0.5\columnwidth]{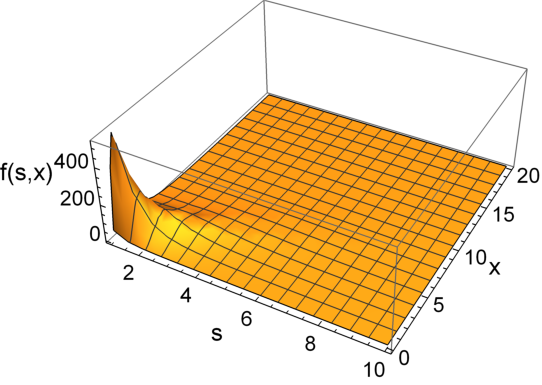}\\
	\caption{ \label{fig1}
    The function $f(s,x)$ for the square configuration,
    with the variable $s$ being the side length of the square.
    (See Fig.\ref{fig8} in Sect 5.)
    Here the $\cos$ mode in $G^{(2)}$ is used for illustration.
         }
\end{figure}

The special solution $\eta^I$ of \eqref{u-equ} is due to the inhomogeneous term,
and can be derived  as the following.
Introduce a Green's function $g(\mathbf{x} , \mathbf{r})$ satisfying
the following equation
\begin{equation} \label{g-equ}
	\big( \nabla^2_\mathbf{x} + 2 k_J^2 \big) g(\mathbf{x}, \mathbf{r})
	= -\delta^{(3)}(\mathbf{x-r}) ,
\end{equation}
and its solution  resembles $G^{(2)}$  of \eqref{gsolution} up to the  constant $\alpha$,
\be
g(\mathbf{x} , \mathbf{r})
= \alpha G^{(2)}(\mathbf{x} , \mathbf{r}) .
\ee
Then  the special solution $\eta^I$
is given by an integration of the product of the Green's function
with the inhomogeneous term (see for instance Ref. \cite{Hackbusch2017})
\bl
\eta^I ({\bf r}_1,{\bf r}_2,{\bf r}_3,{\bf r}_4)
& =\int_V g(\mathbf{x} , \mathbf{r}_1)
\big[2 k_J^2  G^{(2)}({\bf x}, {\bf r}_2) G^{(2)}({\bf x}, {\bf r}_3)
G^{(2)}({\bf x}, {\bf r}_4) \big] \mathrm{d}^3 \mathbf{x} ,
\label{usolution}
\el
where $V$ is an integration volume.
$\eta^I$ is parity-even.
We shall show that $\eta^I$ has a small amplitude.
Write
\bl \label{us}
	\eta^I(z) = \int_{0}^{\infty} f(z,  \mathbf{x})
 \, x^2 d  x ,
\el
where the integrand
\ba \label{fsx}
f(z,  \mathbf{x} )	=\frac{k_J^2}{2 \pi\,   A_\mathrm{m}}
   \int_{0}^{2 \pi} \mathrm{d} \phi 	\int_0^\pi
G^{(2)}({\bf x}, {\bf r}_1) G^{(2)}({\bf x}, {\bf r}_2)
G^{(2)}({\bf x}, {\bf r}_3) G^{(2)}({\bf x}, {\bf r}_4)
 \sin \theta   \mathrm{d} \theta ,
\ea
and $z$ denotes  $({\bf r}_1,{\bf r}_2,{\bf r}_3,{\bf r}_4)$.
For a special case of the square configuration,
we plot the integrand $ f(s, \bf{x})$ in Fig.\ref{fig1}
where  the variable $s$ is the side length of the square.
One sees that $f(s, \bf{x})$ is low in amplitude
except within a small neighborhood around the origin.
Integration confirms that $\eta^I$ is small  (see Sect. 5).

The  homogeneous equation corresponding to Eq.\eqref{u-equ} is
\be  \label{u-equ0}
\big( \nabla^2_\mathbf{r}  + 2 k_J^2 \big) \eta^0 ({\bf r}_1,{\bf r}_2,{\bf r}_3,{\bf r}_4)
= 0  ,
\ee
and a general,   homogeneous  solution is
\be\label{solvTrial}
\eta^0\left(\mathbf{r}_1, \mathbf{r}_2, \mathbf{r}_3, \mathbf{r}_4\right)
=y\left(\mathbf{r}_1\right) y\left(\mathbf{r}_2\right)y\left(\mathbf{r}_3\right)
y\left(\mathbf{r}_4\right) ,
\ee
where each factor function  $y(\bf r)$ is similar to \eqref{G3y},
in which we can set the coefficient $d_l=-1$ without loss of generality.
The homogeneous  solution \eqref{solvTrial}
holds in a general spherical coordinate system and is symmetric under permutations.
To compare with the observation \cite{Philcox2021, Philcox2022,Hou2022},
we can choose the origin of the coordinate system at ${\mathbf r}_4=0$.
(This choice is allowed by the assumption of statistical homogeneity
and also adopted in the observational data.)
So  $y({\bf r}_4)$ becomes a constant and
the homogeneous solution (\ref{solvTrial}) is written as
\bl \label{solvTrial2}
\eta^0\left(\mathbf{r}_1, \mathbf{r}_2, \mathbf{r}_3, \mathbf 0 \right)
& = y\left(\mathbf{r}_1\right) y\left(\mathbf{r}_2\right)y\left(\mathbf{r}_3\right)
\nn \\
& = \sum_{l_1, l_2, l_3} \,  \sum_{m_1, m_2, m_3} C_{l_1 m_1 } C_{l_2 m_2 } C_{l_3 m_3}
Y_{l_1}^{m_1} ( \hat {\bf r}_1 )  Y_{l_2}^{m_2} ( \hat {\bf r}_2 ) Y_{l_3}^{m_3} ( \hat {\bf r}_3 )
   \Big[ c_{l_1}  j_{l_1} (k_0 |\mathbf{r}_1|) - n_{l_1} (k_0 |\mathbf{r}_1|) \Big]
\nn \\
 & ~~~~  \times
  \Big[ c_{l_2}  j_{l_2}  (k_0 |\mathbf{r}_2|) - n_{l_2} (k_0 |\mathbf{r}_2| ) \Big]
 \Big[  c_{l_3}  j_{l_3}  (k_0 |\mathbf{r}_3|) - n_{l_3} (k_0 |\mathbf{r}_3| ) \Big]  ,
\el
where the coefficients $C_{l_i m_j }$, $c_{l_i}$ are arbitrary constants,
and the  summations  are
over $l_i=0,1,...,$ and $m_i=-l_i,..., l_i$ for $i=1,2,3$.
The solution \eqref{solvTrial2} is not yet explicitly isotropic,
ie, invariant under simultaneous rotations of the vectors
$\hat{\mathbf{r}}_1, \hat{\mathbf{r}}_2, \hat{\mathbf{r}}_3$.
To be concordant
with the observation data   \cite{Philcox2022,Philcox2021,Hou2022},
we adopt the isotropic basis functions
 \cite{CahnSlepian2020} for the angular sector,
\bl \label{defbasis}
\mathcal{P}_{l_1  l_2  l_3}
     (\hat{\mathbf{r}}_1, \hat{\mathbf{r}}_2, \hat{\mathbf{r}}_3)
= (-1)^{l_1+ l_2 + l_3} \sum_{m_1, m_2, m_3}
\left(
\begin{array}{c}
 l_1 ~~\, l_2 ~~\, l_3 \\
 m_1 \, m_2 \, m_3
\end{array}
\right)
Y^{m_1}_{l_1} (\hat{\bf  r}_1)
Y^{m_2}_{l_2} (\hat{\bf  r}_2)
Y^{m_3}_{l_3} (\hat{\bf  r}_3) ,
\el
where   the $3 \times 2$ matrix  is a Wigner $3-j$ symbol,
satisfies the triangle condition $|l_1 -l_2| \leq l_3 \leq l_1+l_2$,
and $m_1= -l_1, ... ,  l_1$, etc.
The set of $\mathcal{P}_{l_1  l_2  l_3}$
form a complete orthonormal basis for any isotropic function of
$( {\bf \hat{r}}_1, {\bf \hat{r}}_2,{\bf \hat{r}}_3)$.
Under the parity operation $\mathbb{P}$,
 ${\bf r}_i \rightarrow {- \bf r}_i $,
$  (\theta , \phi) \rightarrow  ( \pi- \theta , \phi+\pi )$,
the basis function transforms as
\bl
  \mathbb{P} \, [ \mathcal{P}_{l_1  l_2  l_3} (\hat{\mathbf{r}}_1, \hat{\mathbf{r}}_2, \hat{\mathbf{r}}_3)]
& =  \mathcal{P}_{l_1  l_2  l_3} (-\hat{\mathbf{r}}_1, -\hat{\mathbf{r}}_2, -\hat{\mathbf{r}}_3)
\nn \\
& = (-1)^{l_1+ l_2 + l_3}
 \mathcal{P}_{l_1  l_2  l_3} (\hat{\mathbf{r}}_1, \hat{\mathbf{r}}_2, \hat{\mathbf{r}}_3),
\el
and is parity-odd and imaginary for $l_1 + l_2 + l_3=$ odd,
 and parity-even and real for $l_1 + l_2 + l_3=$  even.
The  homogeneous solution \eqref{solvTrial2} can be written
in terms of the basis functions as  the following
\bl \label{solv3}
\eta^0\left(\mathbf{r}_1, \mathbf{r}_2, \mathbf{r}_3, \mathbf 0 \right)
& = \sum_{l_1, l_2, l_3} C_{l_1\,  l_2 \, l_3}
  \mathcal{P}_{l_1  l_2  l_3} (\hat{\mathbf{r}}_1, \hat{\mathbf{r}}_2, \hat{\mathbf{r}}_3)
 \Big[ c_{l_1}  j_{l_1} (k_0 |\mathbf{r}_1|) - n_{l_1} (k_0 |\mathbf{r}_1|) \Big]
\nn \\
& ~~~~ \times
  \Big[ c_{l_2}  j_{l_2} (k_0 |\mathbf{r}_2|) - n_{l_2} (k_0 |\mathbf{r}_2|) \Big]
 \Big[  c_{l_3} j_{l_3} (k_0 |\mathbf{r}_3|) - n_{l_3}  (k_0 |\mathbf{r}_3|) \Big]     ,
\el
with
\bl
C_{l_1\,  l_2 \, l_3} \equiv  (-1)^{l_1+ l_2 + l_3}
\sum_{ m_1, m_2, m_3} C_{l_1 m_1 } C_{l_2 m_2 } C_{l_3 m_3} \,
\left(
\begin{array}{c}
 l_1 ~~\, l_2 ~~\, l_3 \\
 m_1 \, m_2 \, m_3
\end{array}
\right)^{-1} ,
\el
where $\left(
\begin{array}{c}
 l_1 ~~\, l_2 ~~\, l_3 \\
 m_1 \, m_2 \, m_3
\end{array}
\right)^{-1}$ is the reciprocal of the Wigner $3-j$ symbol.
Now the solution  \eqref{solv3} is explicitly isotropic.
The amplitude of the solution \eqref{solv3}
is largely specified by the coefficients $C_{l_1\,  l_2 \, l_3}$,
which will be determined by the boundary conditions.
Radially, the homogenous solution \eqref{solv3} is a combination of
two types of modes, $n_l$ and $j_l$,
and generally exhibits an oscillatory characteristic,
which is determined by the reduced Jeans wavenumber $k_0$.
To compare with the observation data,
the homogeneous solution \eqref{solv3}
can be split into the parity-even and parity-odd parts,
\bl \label{solv3even}
\eta^0_{even}\left(\mathbf{r}_1, \mathbf{r}_2, \mathbf{r}_3, \mathbf 0 \right)
& = \sum_{l_1+ l_2+ l_3=even} C_{l_1\,  l_2 \, l_3}
  \mathcal{P}_{l_1  l_2  l_3} (\hat{\mathbf{r}}_1, \hat{\mathbf{r}}_2, \hat{\mathbf{r}}_3)
 \Big[ c_{l_1}  j_{l_1} (k_0 |\mathbf{r}_1|) - n_{l_1} (k_0 |\mathbf{r}_1|) \Big]
\nn \\
 & \times
  \Big[ c_{l_2}  j_{l_2} (k_0 |\mathbf{r}_2|) - n_{l_2} (k_0 |\mathbf{r}_2|) \Big]
 \Big[ c_{l_3} j_{l_3} (k_0 |\mathbf{r}_3|) - n_{l_3}  (k_0 |\mathbf{r}_3|) \Big] ,
\el
\bl \label{solv3odd}
\eta^0_{odd}\left(\mathbf{r}_1, \mathbf{r}_2, \mathbf{r}_3, \mathbf 0 \right)
& = \sum_{l_1+ l_2+ l_3=odd} C_{l_1\,  l_2 \, l_3}
  \mathcal{P}_{l_1  l_2  l_3} (\hat{\mathbf{r}}_1, \hat{\mathbf{r}}_2, \hat{\mathbf{r}}_3)
 \Big[ c_{l_1}  j_{l_1} (k_0 |\mathbf{r}_1|) - n_{l_1} (k_0 |\mathbf{r}_1|) \Big]
\nn \\
 & \times
  \Big[ c_{l_2}  j_{l_2} (k_0 |\mathbf{r}_2|) - n_{l_2} (k_0 |\mathbf{r}_2|) \Big]
 \Big[ c_{l_3} j_{l_3} (k_0 |\mathbf{r}_3|) - n_{l_3}  (k_0 |\mathbf{r}_3|) \Big] .
\el
The 4PCF is real by definition,
whereas $\mathcal{P}_{l_1  l_2  l_3}$ is imaginary for $l_1+ l_2+ l_3=$ odd,
so we can take the imaginary part Im($\eta^0_{odd}$) to represent $\eta^0_{odd}$.

Although convenient for observation,
the homogeneous solution \eqref{solv3} does not explicitly display
the disconnected 4PCF of a Gaussian random process
that is formed from products of two 2PCFs,
and subtracted from the presentation of the observational data
\cite{Philcox2021,Philcox2022}.
For  that, we shall give the following construction.
A  parity-even 4PCF solution of the equation  \eqref{u-equ0}
can be also constructed as a function of ${\bf r}_i-{\bf r}_j$ for $i,j=1,2,3,4$.
This will have explicitly the symmetry under spatial translation
that is not explicit in the solution \eqref{solvTrial}.
Consider a function  $ y({\bf r, r'})$ satisfies
the following homogeneous  equation
\be \label{yeq}
\big( \nabla^2 + 2 k_J^2 \big)  y({\bf r, r'}) =0 ,
\ee
and has a general solution depending on the vector  $({\bf r - r'})$
\bl
y ({\bf r, r'}) & = \sum_{l=0}^\infty  \sum_{m=-l}^{l}
C_{ml}  Y_l^m ( \theta, \phi)
\big[ d_l n_l (k_0 |{\bf r- r'}|)
+ c_l   j_l (k_0 |{\bf r- r'}|) \big] ,
\label{soly}
\el
which is decomposed into  $y=y_{even}+y_{odd}$
with the parity-odd and parity-even parts
\bl
y_{even} ({\bf r, r'}) & = \sum_{l=even}^\infty  \sum_{m=-l}^{l}
C_{ml}  Y_l^m ( \theta, \phi)
\big[ d_l n_l (k_0 |{\bf r- r'}|)
+ c_l   j_l (k_0 |{\bf r- r'}|) \big]  ,
\\
y_{odd} ({\bf r, r'}) & = \sum_{l=odd}^\infty  \sum_{m=-l}^{l}
C_{ml}  Y_l^m ( \theta, \phi)
\big[ d_l n_l (k_0 |{\bf r- r'}|)
+ c_l   j_l (k_0 |{\bf r- r'}|) \big]  .
\el
Using $y_{even}$,
we construct a parity-even 4PCF  homogeneous solution as the following
\bl \label{eta0even}
\eta_{even}(\mathbf{r}_1, \mathbf{r}_2, \mathbf{r}_3, \mathbf{r}_4)
\equiv  & \,  y_{even} (\mathbf{r}_1, \mathbf{r}_2) y_{even}(\mathbf{r}_3, \mathbf{r}_4)
+ y_{even}(\mathbf{r}_1, \mathbf{r}_3) y_{even}(\mathbf{r}_2, \mathbf{r}_4)
\nn \\
& + y_{even}(\mathbf{r}_1, \mathbf{r}_4) y_{even}(\mathbf{r}_2, \mathbf{r}_3) ,
\el
which satisfies the homogeneous equation \eqref{u-equ0}.
(The parity-even \eqref{eta0even}  can be also constructed
out of the general  solution \eqref{solvTrial}
by use of the addition theorem of spherical waves \cite{FriedmanRussek1954,Miller1964}.)
\eqref{eta0even} is symmetric under permutations, as each  $y_{even}$ is even.
In particular, we are interested in
the  $l=0$ term in the even solution \eqref{eta0even},
which is   the following
\bl  \label{disconn}
\eta^{disc} (\mathbf{r}_1, \mathbf{r}_2, \mathbf{r}_3, \mathbf{r}_4)
&  \equiv  \big[ G^{(2)}(r_{12}) G^{(2)} (r_{34})
+ G^{(2)} (r_{13}) G^{(2)} (r_{24})
+ G^{(2)} (r_{14}) G^{(2)} (r_{23}) \big] ,
\el
where $G^{(2)}(r_{12})$ is the 2PCF given by  \eqref{gsolution}.
$\eta^{disc} $ given by \eqref{disconn} is
only a piece of $\eta_{even}$,
referred to as the  disconnected 4PCF
in Refs.\cite{Philcox2021,Philcox2022},
and is equal to the 4PCF of a Gaussian random process,
according to  the Isserlis-Wick theorem \cite{Hristopulos,Isserlis1918,Wick1950}.
All the parts of the 4PCF other than $\eta^{disc}$
are beyond the Gaussian random process.
Again, we see that the self-gravity density fluctuation in the Gaussian approximation
is not a Gaussian random process,
a property also revealed by the nonvanishing 3PCF in Sect 3.
Later we shall display the behavior of the disconnected $\eta^{disc}$
in graphs  in Sect 5.

Note that the parity-odd part of $\eta^0$
can not be constructed in a form analogous to  \eqref{eta0even}
as a sum of products of two $y$'s.
In trying this kind of construction,
the function $y_{odd}({\bf r}_i, {\bf r}_j)$ will be needed,
which nevertheless  violates the permutation symmetry.

In summary,
 the  analytical solution of 4PCF in the Gaussian approximation is given by
\ba  \label{g4solution}
\eta  \equiv \eta^0_{odd} + \eta^0_{even} + \eta^I  +  \eta^{FP}  ,
\ea
where $\eta^0_{odd} + \eta^0_{even}$ is the homogeneous solution
and $\eta^I  +  \eta^{FP}$ is the inhomogeneous solution of \eqref{solve4PCF-1}.
Only $\eta^0_{odd}$ is parity-odd,
$\eta^0_{even}$ is parity-even
which contains the disconnected piece $\eta^{disc}$.
Both $\eta^{FP}$ and $\eta^I$ are parity-even
and are constructed from $G^{(2)}$.

\section{The   4PCF solution compared with observations }

\begin{figure}[htbp]
	\centering
	\includegraphics[width=0.9\columnwidth]{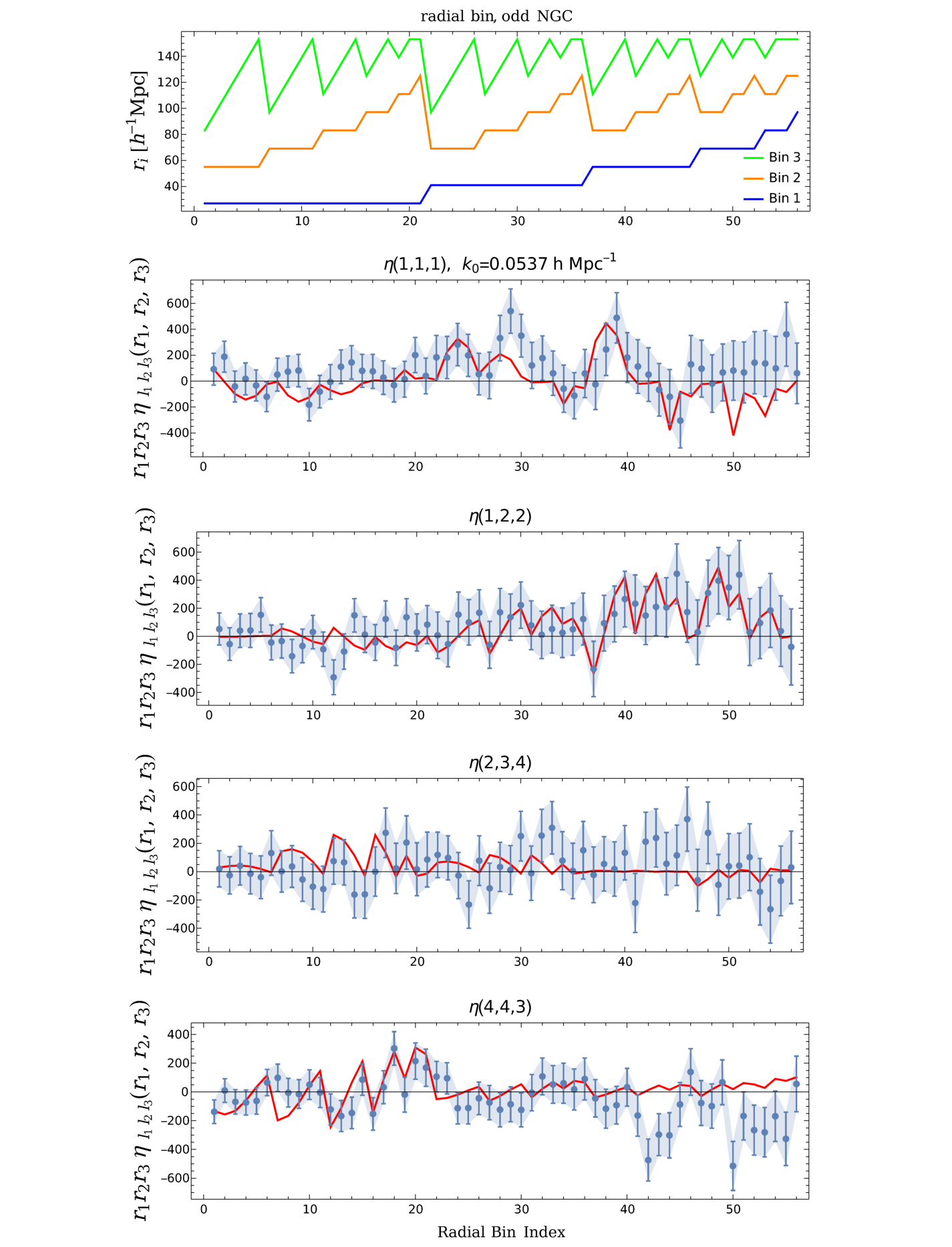}\\
	\caption{  \label{fig2}
	 $\eta^0_{odd}$   (red line)
and the 4PCF data (dots) of NGC  (Northern galactic cap)
    BOSS CMASS from Ref.\cite{Philcox2022}.
With $k_0 = 0.0537 h$ Mpc$^{-1}$ ,
$C_{111} = 0.026,  C_{122} = 0.024,  C_{234} = -0.00031,  C_{443} = 0.00003$,
$c_1 = -1.3,     c_2 = -0.9 ,    c_3 = -5.7,   c_4 = 9.9$ .
	}
\end{figure}

\begin{figure}[htbp]
	\centering
	\includegraphics[width=0.9\columnwidth]{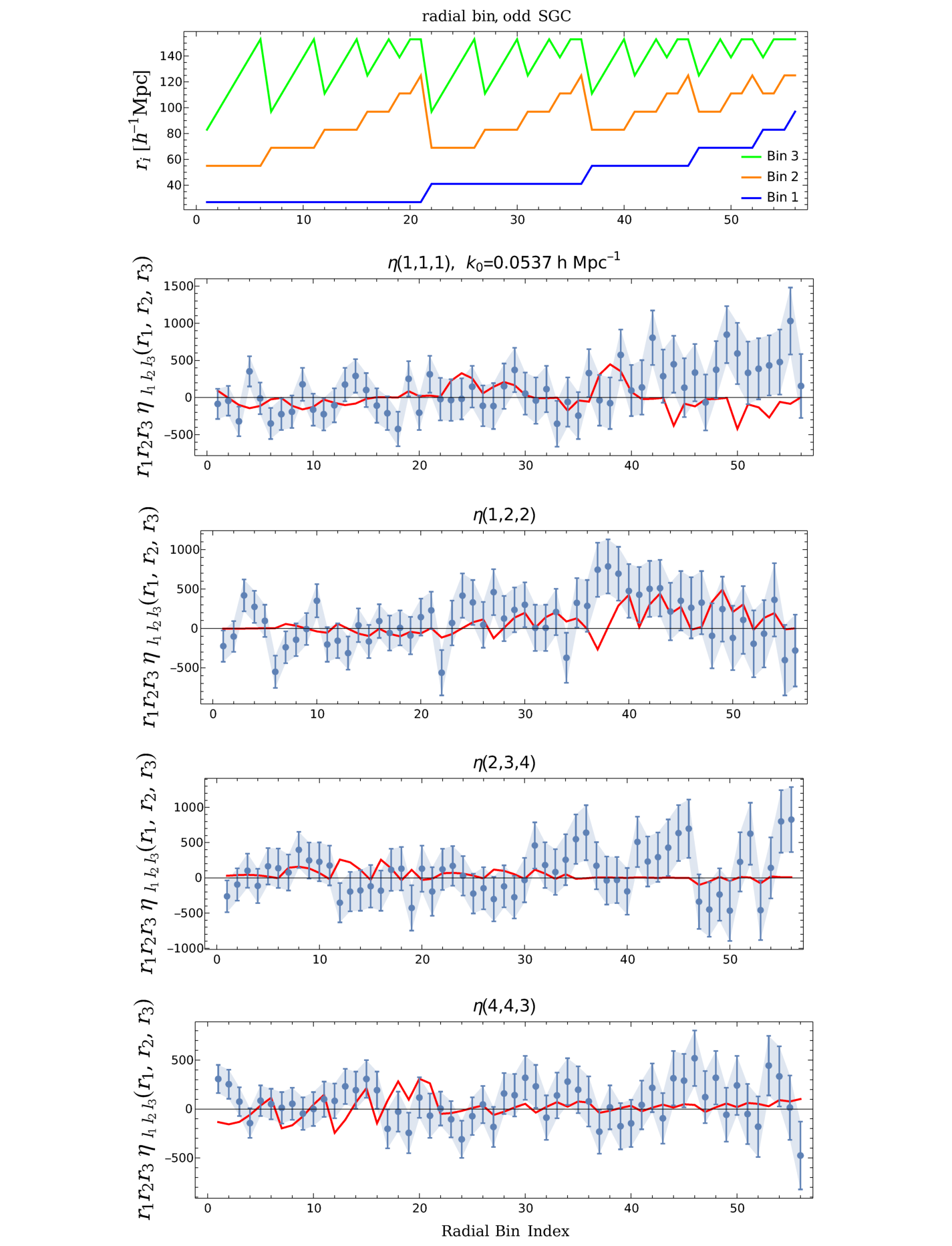}\\
	\caption{  \label{fig3}
	 $\eta^0_{odd}$  (red line)
and the 4PCF data (dots) of SGC (Southern galactic cap)
BOSS CMASS from Ref.\cite{Philcox2022}.
The parameters and the coefficients
are the same as in Fig.\ref{fig2}.
	}
\end{figure}

\newpage

Given the 4PCF solution  \eqref{g4solution},
we shall compare it with the observations in this section.
First consider the parity-odd part $\eta^0_{odd}$.
The observed parity-odd 4PCF from BOSS CMASS
has been given in Fig.2 of Ref. \cite{Philcox2022}
for the four multiplets $(l_1, l_2, l_3)=(1,1,1), (1,2,2), (2,3,4), (4,4,3)$,
each being a function of $r_1,r_2,r_3$.
The theoretical $\eta^0_{odd}$ of \eqref{solv3odd}
is already in the basis $\mathcal{P}_{l_1  l_2  l_3}$
and we just focus on the radial sector.
For each multiplet, we simply set the theoretical = the data,
\bl \label{solv111}
&    C_{111}  \Big[   c_{1}  j_{1} (k_0 |\mathbf{r}_1|) - n_{1} (k_0 |\mathbf{r}_1|) \Big]
   \Big[ c_{1}  j_{1} (k_0 |\mathbf{r}_2|) - n_{1} (k_0 |\mathbf{r}_2|) \Big]
\nn \\
& ~~~~~~~~  \times
 \Big[ c_{1} j_{1} (k_0 |\mathbf{r}_3|) - n_{1}  (k_0 |\mathbf{r}_3|) \Big]
 r_1 r_2 r_3
  =data_{111}(\bf r_1,r_2,r_3)     ,
\\
&    C_{122}  \Big[ c_{1}  j_{1} (k_0 |\mathbf{r}_1|) - n_{1} (k_0 |\mathbf{r}_1|) \Big]
   \Big[ c_{2}  j_{2} (k_0 |\mathbf{r}_2|) - n_{2} (k_0 |\mathbf{r}_2|) \Big]
\nn \\
& ~~~ ~~~~~  \times
 \Big[ c_{2} j_{2} (k_0 |\mathbf{r}_3|) - n_{2}  (k_0 |\mathbf{r}_3|) \Big]
 r_1 r_2 r_3
  =data_{122}(\bf r_1,r_2,r_3)     ,
\label{solv122}
\\
&    C_{234}  \Big[ c_{2}  j_{2} (k_0 |\mathbf{r}_1|) - n_{2} (k_0 |\mathbf{r}_1|) \Big]
   \Big[ c_{3}  j_{3} (k_0 |\mathbf{r}_2|) - n_{3} (k_0 |\mathbf{r}_2|) \Big]
\nn \\
& ~~~ ~~~~~  \times
 \Big[ c_{4} j_{4} (k_0 |\mathbf{r}_3|) - n_{4}  (k_0 |\mathbf{r}_3|) \Big]
 r_1 r_2 r_3
  =data_{234}(\bf r_1,r_2,r_3)     ,
\label{solv234}
\\
&    C_{443}
 \Big[ c_{4}  j_{4} (k_0 |\mathbf{r}_1|) - n_{4} (k_0 |\mathbf{r}_1|) \Big]
   \Big[ c_{4}  j_{4} (k_0 |\mathbf{r}_2|) - n_{4} (k_0 |\mathbf{r}_2|) \Big]
  \nn \\
& ~~~ ~~~~~  \times
 \Big[ c_{3} j_{3} (k_0 |\mathbf{r}_3|) - n_{3}  (k_0 |\mathbf{r}_3|) \Big]
   r_1 r_2 r_3
  =data_{443}(\bf r_1,r_2,r_3)     .
\label{solv443}
\el
and tune the coefficients  $C_{l_1 l_2  l_3}$ and $c_l$.
The same set of parity-odd coefficients $C_{l_1 l_2  l_3}$
and the radial coefficients $c_l$
have been applied to both NGC (Northern galactic cap)
and SGC (Southern galactic cap).
The resulting $\eta^0_{odd}$ and the data are shown
in Fig.\ref{fig2} for NGC, and  in Fig.\ref{fig3} for SGC.
The theoretical $\eta^0_{odd}$ qualitatively explains the observed parity-odd data

\begin{figure}[htbp]
	\centering
	\includegraphics[width=0.9\columnwidth]{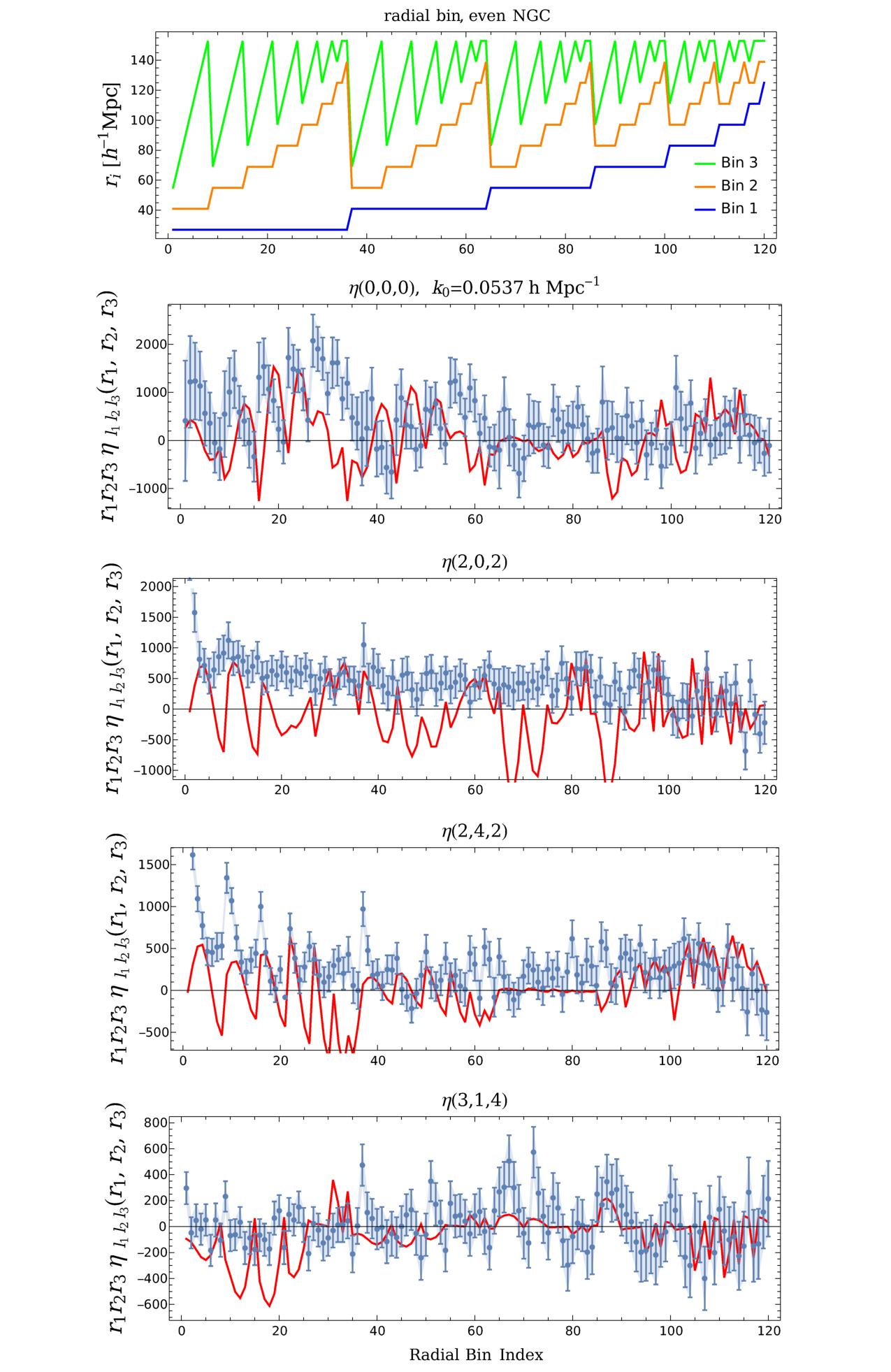}\\
	\caption{  \label{fig4}
	 $\eta^R_{even}$  (red line)
and the 4PCF data (dots) of NGC BOSS CMASS from Ref.\cite{Philcox2021}.
With  $A_m=1.28\, h^{-1}$ Mpc, $d_0=0.65$,
$C_{000} = 0.01,  C_{202} = 0.04, C_{242} = -0.0045,  C_{314} = 0.0008$,
$c_0=-0.53$.
Other parameters and radial coefficients $c_l$
are the same as in Fig.\ref{fig2}
for the parity-odd.
	}
\end{figure}

\begin{figure}[htbp]
	\centering
	\includegraphics[width=0.9\columnwidth]{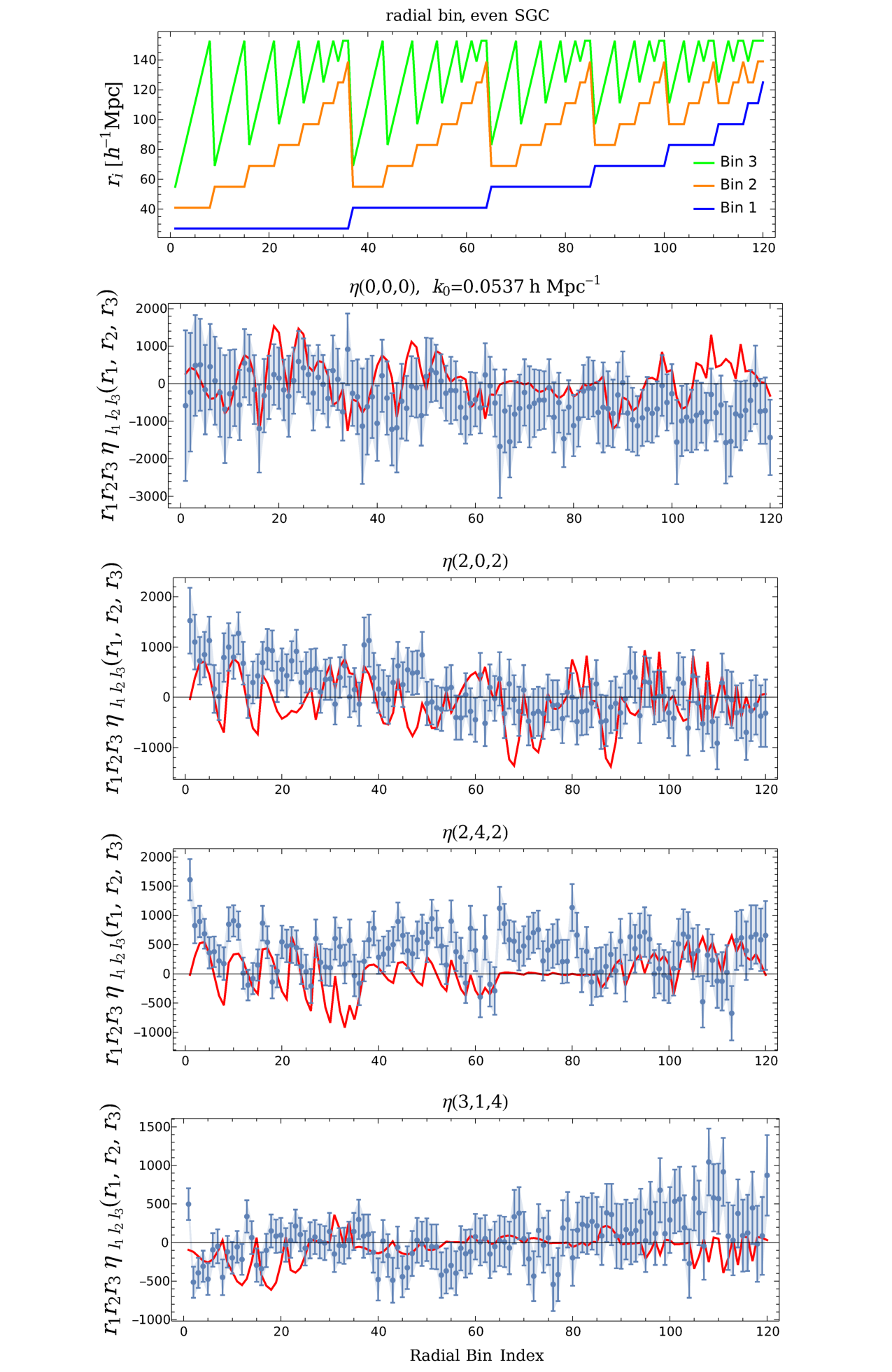}\\
	\caption{  \label{fig5}
	 $\eta^R_{even}$  (red line)
and the 4PCF data (dots) of SGC BOSS CMASS from Ref.\cite{Philcox2021}.
The parameters and the coefficients are the same as Fig.\ref{fig4} for NGC.
	}
\end{figure}

We explain how to understand Fig.\ref{fig2}.
The top plot gives the radial coordinates $r_1,r_2,r_3$
of the 4-point correlation $\eta$ (with  $r_4$ set to $0$),
in which the curve Bin 1 stands for  $r_1$, Bin 2 for $r_2$, and Bin 3  for $r_3$.
Draw a vertical line on the plot.
This vertical line will intersect with the three bin curves in the top plot,
yielding a triple values  $(r_1,r_2,r_3)$ of the coordinates,
simultaneously,
this same vertical line will also intersect with the curve $\eta_{~  l_1,l_2,l_3} $
in the lower panels, yielding a value of  $\eta_{~  l_1,l_2,l_3}(r_1,r_2,r_3)$.
This is an economic way to express the function $\eta_{~  l_1,l_2,l_3}$
with three variables $(r_1,r_2,r_3)$.
As a limitation, the top plot,  as a 2-dim plane,
 does not cover the full range of $(r_1,r_2,r_3)$ in 3-dim space,
but only covers some slices in  it.
The way is the same for Fig.\ref{fig3},  Fig.\ref{fig4} and Fig.\ref{fig5}.

The observational data in Ref.\cite{Philcox2022} shows that
the first two multiplets have higher amplitudes
$C_{111}\sim C_{122}\sim 10^{-2} $,
and  other have lower amplitudes
$C_{234}\sim 10^{-3}$,  $ C_{344}\sim 10^{-4}$.
The radially oscillatory feature in the observed 4PCF
can be explained by the radial modes $n_l$ and $j_l$ in the solution $\eta^0_{odd}$,
and is largely controlled by the Jeans wavenumber $k_J$ of the systems of galaxies.

Next,  consider the parity-even part $\eta^0_{even}+\eta^{FP}+ \eta^I$,
which is more sophiscated than the parity-odd case.
The observed parity-even 4PCF from BOSS CMASS in Fig.3 of Ref.\cite{Philcox2021}
excludes  the disconnected  $\eta^{disc}$ of the Gaussian random process.
To compare with the data,
we subtract off the disconnected  $\eta^{disc}$
to give the reduced parity-even 4PCF as the following
\bl\label{ComputEvenEta}
\eta_{even}^{R} \equiv \eta^0_{even}- \eta^{disc} +\eta^I  + \eta^{FP} ,
\el
where the superscript ``$R$" denotes the reduced.
We remark that $\eta^I$ and $ \eta^{FP}$ are connected,
nevertheless, $\eta^0_{even}$,
by its construction \eqref{solv3even} or \eqref{eta0even},
is not connected, as the lines joining
the vertices $(\mathbf{r}_1, \mathbf{r}_2, \mathbf{r}_3, \mathbf{r}_4)$
are obviously disconnected.
Therefore, $(\eta^0_{even}- \eta^{disc})$ in \eqref{ComputEvenEta}
is not connected.
For this reason we refer to \eqref{ComputEvenEta}
as the ``reduced" even-parity 4PCF,
instead of  the ``connected" even-parity 4PCF in Ref.\cite{Philcox2022}.
Moreover, $\eta^0_{even}$ is angle-dependent
and contains important information of the system of galaxies
beyond $\eta^{disc}$ of a Gaussian random process.
Since  $\eta^I$ is comparatively small,
we can neglect it in a preliminary computing,
\bl\label{EvenEta2}
\eta_{even}^{R} \simeq  \eta^0_{even}- \eta^{disc}  + \eta^{FP} ,
\el
where  $\eta^0_{even}$ is given by \eqref{solv3even}.
Since the data are in the basis $\mathcal{P}_{l_1 l_2 l_3}$,
we need to express $\eta^{disc}$ of \eqref{disconn} and $\eta^{FP}$ of \eqref{solh}
in terms of $\mathcal{P}_{l_1 l_2 l_3}$, too.
As before, taking  $\mathbf{r}_4=0$  leads to
\bl\label{eta0dis}
\eta^{disc} (\mathbf{r}_1, \mathbf{r}_2, \mathbf{r}_3, 0)
& =  \big[ G^{(2)} (r_{12}) G^{(2)} (r_{3})
+ G^{(2)}(r_{13}) G^{(2)}(r_{2})
+  G^{(2)}(r_{23}) G^{(2)}(r_{1}) \big] ,
\el
\bl
\eta^{FP} ({\bf r}_1,{\bf r}_2,{\bf r}_3,0)
 = & \Big[  G^{(2)}(r_{12}) G^{(2)}(r_{23}) +  G^{(2)}(r_{23}) G^{(2)}(r_{31})
   +  G^{(2)}(r_{31}) G^{(2)}(r_{12}) \Big]
   \nn \\
&  ~~~~~  \times \Big[  G^{(2)}(r_{1})+   G^{(2)}(r_{2}) +  G^{(2)}(r_{3}) \Big]
  \nn \\
&  +  G^{(2)}(r_{23})  \Big[ G^{(2)}(r_{1}) G^{(2)}(r_{2})
   +   G^{(2)}(r_{3}) G^{(2)}(r_{1})  \Big]
 \nn \\
& + G^{(2)}(r_{21})  \Big[ G^{(2)}(r_{3}) G^{(2)}(r_{2})
  +   G^{(2)}(r_{1})G^{(2)}(r_{3})   \Big]
 \nn  \\
&  + G^{(2)}(r_{13})  \Big[ G^{(2)}(r_{2}) G^{(2)}(r_{1})
  +    G^{(2)}(r_{2}) G^{(2)}(r_{3})   \Big]
 \nn \\
& +  G^{(2)}(r_{2}) G^{(2)}(r_{3}) G^{(2)}(r_{1}) ,
\label{etaFPr}
\el
both being functions  of six variables
$r_1$, $r_2$, $r_3$, $r_{12}$, $r_{13}$, $r_{23}$.
By statistical isotropy, we can take the vector ${\bf r}_3$ along the $z$-axis,
the vector ${\bf r}_2$ on the $x-z$-plane,
so that $\mathbf{r}_2=(r_2, \theta_2, \phi_2=0)$,
 $\mathbf{r}_1=(r_1, \theta_1, \phi_1)$.
Then
\bl
r_{12} &= \sqrt{r_1^2 + r_2^2 -2 {\bf r_1 \cdot r_2} }
\nn \\
& = \sqrt{r_1^2 + r_2^2 -2 r_1 r_2 (\sin \theta_1 \cos\phi_1 \sin \theta_2 \cos\phi_2
+ \sin \theta_1 \sin \phi_1 \sin \theta_2 \sin \phi_2
+ \cos \theta_1  \cos \theta_2 ) }
\nn \\
& = \sqrt{r_1^2 + r_2^2 -2 r_1 r_2 (\sin \theta_1 \cos\phi_1 \sin \theta_2
+ \cos \theta_1  \cos \theta_2 ) }   ,
\nn \\
r_{13} & = \sqrt{r_1^2 + r_3^2 -2 {\bf r_1 \cdot r_3} }
= \sqrt{r_1^2 + r_3^2 -2 r_1 r_3 \cos \theta_1} \,  ,
\nn \\
r_{23} & = \sqrt{r_2^2 + r_3^2 -2 {\bf r_2 \cdot r_3} }
=  \sqrt{r_2^2 + r_3^2 -2 r_2 r_3 \cos \theta_2 } \, ,
 \label{etaevenrrr}
\el
which are expressed in terms the three angular variables $\phi_1,  \theta_1, \theta_2$.
Now we project \eqref{eta0dis} and \eqref{etaFPr},   respectively,
into the basis $\mathcal{P}_{l_1  l_2  l_3}$,
\bl
\eta^{disc} (\mathbf{r}_1, \mathbf{r}_2, \mathbf{r}_3, 0)
=   \sum_{l_1, l_2, l_3}   \eta^{disc}_{l_1 l_2 l_3}(r_1, r_2, r_3)
  \mathcal{P}_{l_1 l_2 l_3}
   (\hat{\mathbf{r}}_1, \hat{\mathbf{r}}_2, \hat{\mathbf{r}}_3) ,
   \\
\eta^{FP} (\mathbf{r}_1, \mathbf{r}_2, \mathbf{r}_3, 0)
=   \sum_{l_1, l_2, l_3}   \eta^{FP}_{l_1 l_2 l_3}(r_1, r_2, r_3)
  \mathcal{P}_{l_1 l_2 l_3}
   (\hat{\mathbf{r}}_1, \hat{\mathbf{r}}_2, \hat{\mathbf{r}}_3) ,
\el
where the multiplets are given by the inner product with the basis
\bl \label{etaevenlll}
\eta^{disc}_{l_1 l_2 l_3} (r_1, r_2, r_3)=\int d \hat{\mathbf{r}}_1
  d \hat{\mathbf{r}}_2 d \hat{\mathbf{r}}_3
  \eta^{disc} (\mathbf{r}_1, \mathbf{r}_2, \mathbf{r}_3)
  \mathcal{P}^*_{l_1 l_2 l_3}(\hat{\mathbf{r}}_1, \hat{\mathbf{r}}_2, \hat{\mathbf{r}}_3) ,
\\
\eta^{FP}_{l_1 l_2 l_3} (r_1, r_2, r_3)=\int d \hat{\mathbf{r}}_1
  d \hat{\mathbf{r}}_2 d \hat{\mathbf{r}}_3
  \eta^{FP} (\mathbf{r}_1, \mathbf{r}_2, \mathbf{r}_3)
  \mathcal{P}^*_{l_1 l_2 l_3}(\hat{\mathbf{r}}_1, \hat{\mathbf{r}}_2, \hat{\mathbf{r}}_3) .
\el
By the prescribed orientation,  we set
 $Y_l^m (\hat{\mathbf{r}}_3)=1$ and $\phi_2=0$ in the basis function
$\mathcal{P}^*_{l_1 l_2 l_3}(\hat{\mathbf{r}}_1, \hat{\mathbf{r}}_2, \hat{\mathbf{r}}_3)$,
and $\int d \hat{\mathbf{r}}_3=4\pi$, $\int d\phi_2 =2\pi $.
Four parity-even multiplets, $(0,0,0)$, $(2,0,2)$, $(2,4,2)$, $(3,1,4)$,
have been given in the data \cite{Philcox2021},
so we shall compute the following  three-fold angular integrations
\bl
\eta^{disc}_{000} (r_1, r_2, r_3)
&  =8 \pi^2 \int \sin\theta_1 d\theta_1 d\phi_1  \sin\theta_2  d \theta_2
 \eta^{disc} (\mathbf{r}_1, \mathbf{r}_2, |r_3|)  \mathcal{P}^*_{000} \,  ,
 \\
\eta^{disc}_{202} (r_1, r_2, r_3)
& =8 \pi^2 \int \sin\theta_1 d\theta_1 d\phi_1  \sin\theta_2  d \theta_2
 \eta^{disc} (\mathbf{r}_1, \mathbf{r}_2, r_3)
  \mathcal{P}^*_{202}(\hat{\mathbf{r}}_1, \hat{\mathbf{r}}_2, \hat{z} )  ,
 \\
\eta^{disc}_{242} (r_1, r_2, r_3)
& = 8 \pi^2 \int \sin\theta_1 d\theta_1 d\phi_1  \sin\theta_2  d \theta_2
 \eta^{disc} (\mathbf{r}_1, \mathbf{r}_2, r_3)
  \mathcal{P}^*_{242}(\hat{\mathbf{r}}_1, \hat{\mathbf{r}}_2, \hat{ z})   ,
   \\
\eta^{disc}_{314} (r_1, r_2, r_3)
& = 8 \pi^2 \int \sin\theta_1 d\theta_1 d\phi_1  \sin\theta_2  d \theta_2
 \eta^{disc} (\mathbf{r}_1, \mathbf{r}_2, r_3)
  \mathcal{P}^*_{314}(\hat{\mathbf{r}}_1, \hat{\mathbf{r}}_2, \hat z)   ,
\el
\bl
\eta^{FP}_{000} (r_1, r_2, r_3)
&  =8 \pi^2 \int \sin\theta_1 d\theta_1 d\phi_1  \sin\theta_2  d \theta_2
         \,   \eta^{FP} (\mathbf{r}_1, \mathbf{r}_2, |r_3|)
         \mathcal{P}^*_{000}  ,
         \\
\eta^{FP}_{202} (r_1, r_2, r_3)
& =8 \pi^2 \int \sin\theta_1 d\theta_1 d\phi_1  \sin\theta_2  d \theta_2
 \eta^{FP} (\mathbf{r}_1, \mathbf{r}_2, r_3)
  \mathcal{P}^*_{202}(\hat{\mathbf{r}}_1, \hat{\mathbf{r}}_2, \hat{z} ) ,
 \\
\eta^{FP}_{242} (r_1, r_2, r_3)
& = 8 \pi^2 \int \sin\theta_1 d\theta_1 d\phi_1  \sin\theta_2  d \theta_2
 \eta^{FP} (\mathbf{r}_1, \mathbf{r}_2, r_3)
  \mathcal{P}^*_{242}(\hat{\mathbf{r}}_1, \hat{\mathbf{r}}_2, \hat{ z}) ,
   \\
\eta^{FP}_{314} (r_1, r_2, r_3)
& = 8 \pi^2 \int \sin\theta_1 d\theta_1 d\phi_1  \sin\theta_2  d \theta_2
 \eta^{FP} (\mathbf{r}_1, \mathbf{r}_2, r_3)
  \mathcal{P}^*_{314}(\hat{\mathbf{r}}_1, \hat{\mathbf{r}}_2, \hat z) .
\el
These integration can be done numerically.
We set the theoretical = the data,
\bl
& C_{000}
\Big[ c_{0}  j_{0} (k_0 |\mathbf{r}_1|) - n_{0} (k_0 |\mathbf{r}_1|) \Big]
\Big[ c_{0}  j_{0} (k_0 |\mathbf{r}_2|) - n_{0} (k_0 |\mathbf{r}_2|) \Big]
\nn \\
& ~~~~~ \times
\Big[ c_{0}  j_{0} (k_0 |\mathbf{r}_3|) - n_{0} (k_0 |\mathbf{r}_3|) \Big]
      r_1 r_2 r_3
+  \Big[ - \eta^{disc}_{000}(r_1, r_2, r_3)+ \eta^{FP}_{000}(r_1, r_2, r_3) \Big] r_1 r_2 r_3
\nn
\\
& ~~~~~    =data_{000}(\bf r_1,r_2,r_3)   ,
  \nn \\
& C_{202}  \Big[  c_{2}  j_{2} (k_0 |\mathbf{r}_1|) - n_{2} (k_0 |\mathbf{r}_1|) \Big]
   \Big[ c_{0}  j_{0} (k_0 |\mathbf{r}_2|) - n_{0} (k_0 |\mathbf{r}_2|) \Big]
\nn \\
& ~~~~~  \times
 \Big[ c_{2} j_{2} (k_0 |\mathbf{r}_3|) - n_{2}  (k_0 |\mathbf{r}_3|) \Big]
 r_1 r_2 r_3
 + \Big[ - \eta^{disc}_{202}(r_1, r_2, r_3) + \eta^{FP}_{202}(r_1, r_2, r_3) \Big] r_1 r_2 r_3
\nn
\\
& ~~~~~   =data_{202}(\bf r_1,r_2,r_3)      ,
  \nn \\
& C_{242}  \Big[  c_{2}  j_{2} (k_0 |\mathbf{r}_1|) - n_{2} (k_0 |\mathbf{r}_1|) \Big]
   \Big[ c_{4}  j_{4} (k_0 |\mathbf{r}_2|) - n_{4} (k_0 |\mathbf{r}_2|) \Big] ,
\nn \\
& ~~~~~  \times
 \Big[ c_{2} j_{2} (k_0 |\mathbf{r}_3|) - n_{2}  (k_0 |\mathbf{r}_3|) \Big]
 r_1 r_2 r_3
  + \Big[ - \eta^{disc}_{242}(r_1, r_2, r_3) + \eta^{FP}_{242}(r_1, r_2, r_3)  \Big] r_1 r_2 r_3
\nn
\\
& ~~~~~  =data_{242}(\bf r_1,r_2,r_3)      ,
  \label{evencomp242}
  \\
& C_{314}  \Big[ c_{3}  j_{3} (k_0 |\mathbf{r}_1|) - n_{3} (k_0 |\mathbf{r}_1|) \Big]
   \Big[ c_{1}  j_{1} (k_0 |\mathbf{r}_2|) - n_{1} (k_0 |\mathbf{r}_2|) \Big]
\nn \\
& ~~~~~ \times
 \Big[ c_{4} j_{4} (k_0 |\mathbf{r}_3|) - n_{4}  (k_0 |\mathbf{r}_3|) \Big]
 r_1 r_2 r_3
  + \Big[ - \eta^{disc}_{314}(r_1, r_2, r_3)
  + \eta^{FP}_{314}(r_1, r_2, r_3)  \Big] r_1 r_2 r_3
\nn
\\
&  ~~~~~  = data_{314}(\bf r_1,r_2,r_3)      ,
  \label{evencomp314}
\el
and tune the parity-even coefficients $C_{l_1 l_2  l_3}$.
The radial coefficients $c_1,c_2,c_3,c_4$
are shared by the parity-even and the parity-odd.
The resulting $\eta^R_{even}$
and the data are shown in Fig.\ref{fig4} for NGC,
and respectively in Fig.\ref{fig5} for SGC.
The observed parity-even 4PCF
has a higher amplitude than the observed parity-odd 4PCF  by several times.

Comparatively, the fitting for
the parity-odd in Fig.\ref{fig2} and Fig.\ref{fig3} seems better than
that for the parity-even in Fig.\ref{fig4} and Fig.\ref{fig5}.
This may be due to our simple numerical computing of
the integrations of  $\eta^{disc}$ and $\eta^{FP}$,
or due to the absence of the nonlinear terms in the Gaussian approximation.
Overall, the fitting in this paper is still preliminary,
and could be improved by the parameter estimation
which would involve more than fifteen parameters.

\newpage

\section{The behavior of some parts of the parity-even 4PCF}

The parity-even 4PCF is composed of three pieces $\eta^0_{even}+\eta^{FP}+\eta^I$,
and we like to compare their respective contribution.
The piece $\eta^0_{even}$ has the angular sector and is not simple to plot.
For simplicity,  we shall demonstrate the behaviors of
 $\eta^{disc}$, $\eta^{FP}$, and $\eta^I$ respectively.
In the following we consider two special configurations:
a conformal line and a square that are adopted in Ref.\cite{FryPeebles}.

\begin{figure}[htb]
	\centering
	\includegraphics[width = 0.4\linewidth]{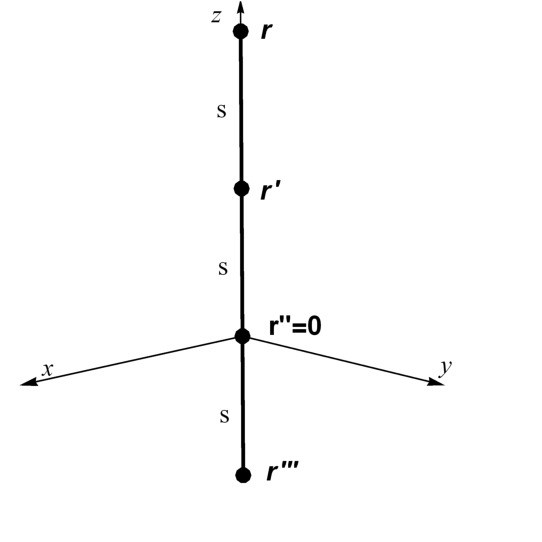}
	\caption{ Four galaxies are located along the $z$-axis
    with an equal separation  $s$ between neighboring galaxies.
	The origin is at $\mathbf{r''=0}$.
	}
	\label{fig6}
\end{figure}
\begin{figure}[htbp]
	\centering
	\includegraphics[width=0.8\columnwidth]{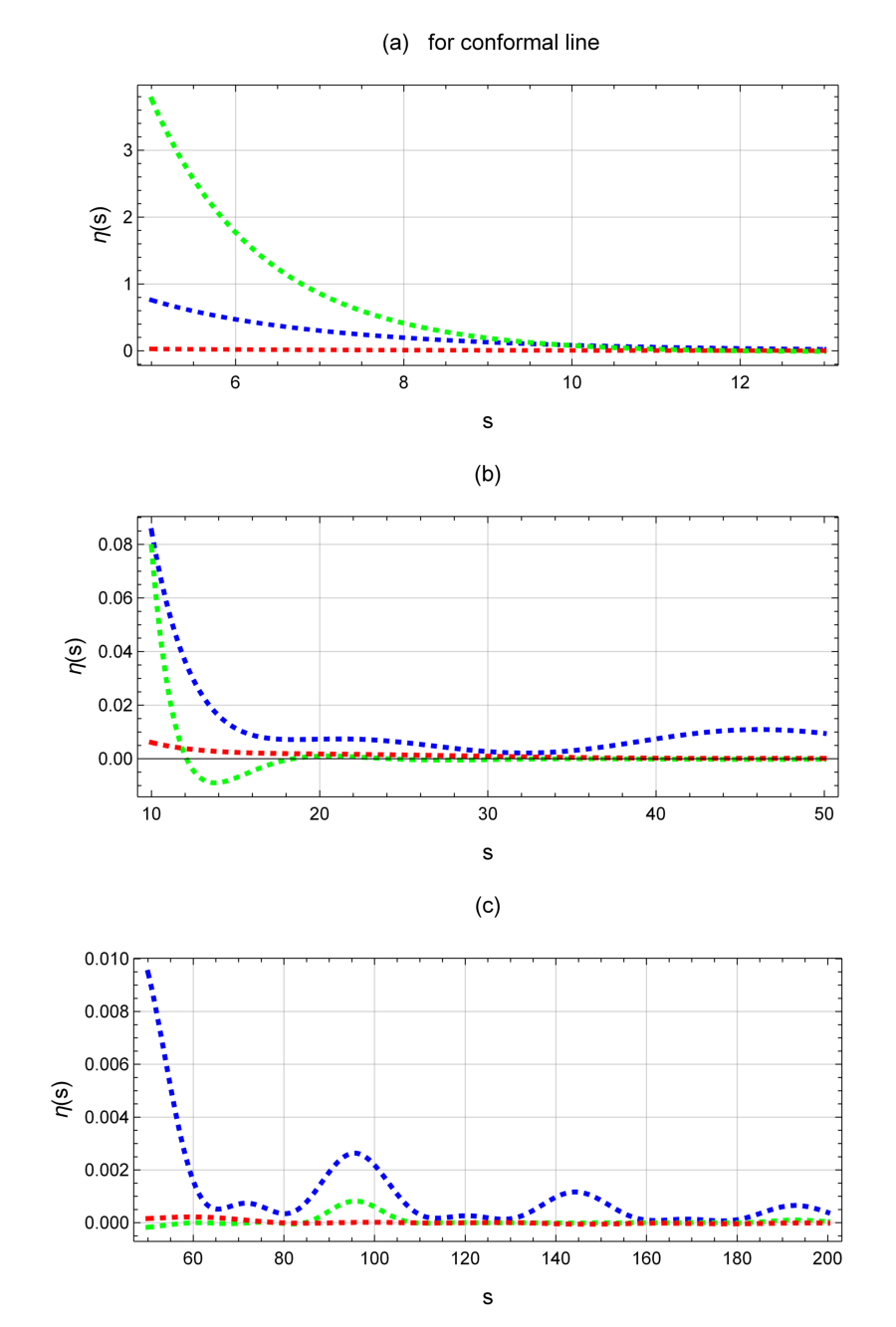}\\
	\caption{ \label{fig7}
The conformal line configuration.
		(a) $\eta^{FP}$ (green) is dominant at small scales
      and decreases with separation.
		(b) $\eta^{disc}$ (blue) starts to dominate at middle  scales.
     (c) $\eta^{disc}$ is oscillatory at large scales.
      $\eta^I$ (red) is small.
     For illustration, the radial mode $\frac{\cos x}{x}$ in $\xi$ is used.
	}
\end{figure}

In the conformal line case,
the four galaxies  with position $(\mathbf{r, r', r'', r'''})$
are put on a line,
and the separation between two neighboring galaxies is  $s$.
In the spherical coordinate $(r, \theta, \phi)$,
we consider the line along the $z$-axis, so that
$\mathbf{r''}=(0, 0, 0)$, $\mathbf{r'''}=(s, \pi, 0)$,
 $\mathbf{r'}=(s, 0, 0)$, and $\mathbf{r}=(2s, 0, 0)$,
as shown in Fig. \ref{fig6}.
Given  the conformal line, one has
\bl
\eta^{disc}(s)  & =   \xi(s)^2 + \xi(2s)^2 + \xi(s) \xi(3s) ,
\label{eta0line}  \\
\eta^{FP}(s) & =  4 \xi(s) \xi(2 s) \xi(3 s)
+4 \xi(s)^2 \xi(2 s)
+3 \xi(s) \xi(2 s)^2
 \\
& ~~~ +3 \xi(s)^2 \xi(3 s) +\xi(2 s)^2 \xi(3 s)  + \xi(s)^3 ,
\nn
\el
and $\eta^I$ is the integration  \eqref{usolution}.
With the integration variable
$\mathbf{x}=(x, \theta, \phi)$ in the spherical coordinate,
it is
\bl
\eta^I(s)=&\frac{k_0^2}{4 \pi A_\mathrm{m}}
\int_V  \xi(\sqrt{x^2+4 s^2-4 s x \cos \theta})
\xi(\sqrt{x^2+ s^2-2 s x \cos \theta }) \nn \\
& ~~~~ \times \xi(\sqrt{x^2+ s^2+2 s x \cos \theta})
\xi(x) x^2 \sin \theta \, \mathrm{d} x \mathrm{d} \theta \mathrm{d} \phi .
\el
The results are shown in Fig.\ref{fig7}
for the conformal line configuration.
At small scales  $r \lesssim  10$Mpc
the Fry-Pebbles part $\eta^{FP} \propto \xi^3 \propto r^{-3.2}$  (by an actual fit)
is dominant since $\xi \propto r^{-1}$ in the Gaussian approximation.
At large scales  $r\gtrsim 10$Mpc
the disconnected part $\eta^{disc}  \propto \xi^2 \propto r^{-2}$ is dominant
 and oscillates with a period $\pi/k_0 \sim 60$Mpc.
$\eta^I$ is comparable, or subdominant to $\eta^{FP}$ at large scales.
The detailed of the curves are subject to varying with the parameters $k_0$ and $A_m$.

\newpage

\begin{figure}[htb]
	\centering
	\includegraphics[width = 0.4\linewidth]{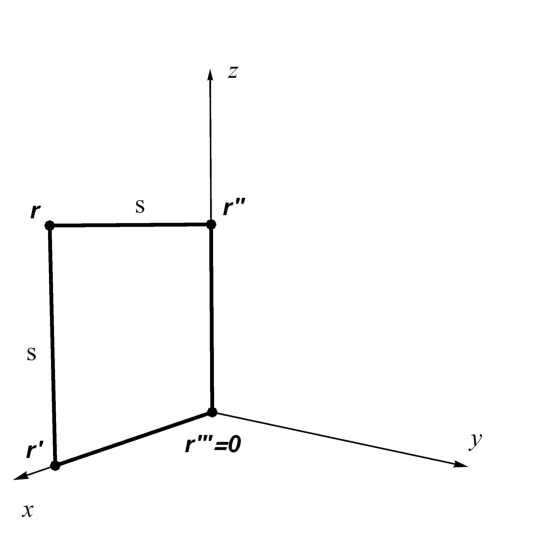}
	\caption{ Four galaxies are located on a square.
The origin	is at  $\mathbf{r'''} =0$,
	$(\mathbf{r'}-\mathbf{r''' })$ is along  the $x$-axis,
and $(\mathbf{r''}-\mathbf{r''' })$ along the $z$-axis, respectively.
	}
	\label{fig8}
\end{figure}

\begin{figure}[htbp]
	\centering
	\includegraphics[width=0.8\columnwidth]{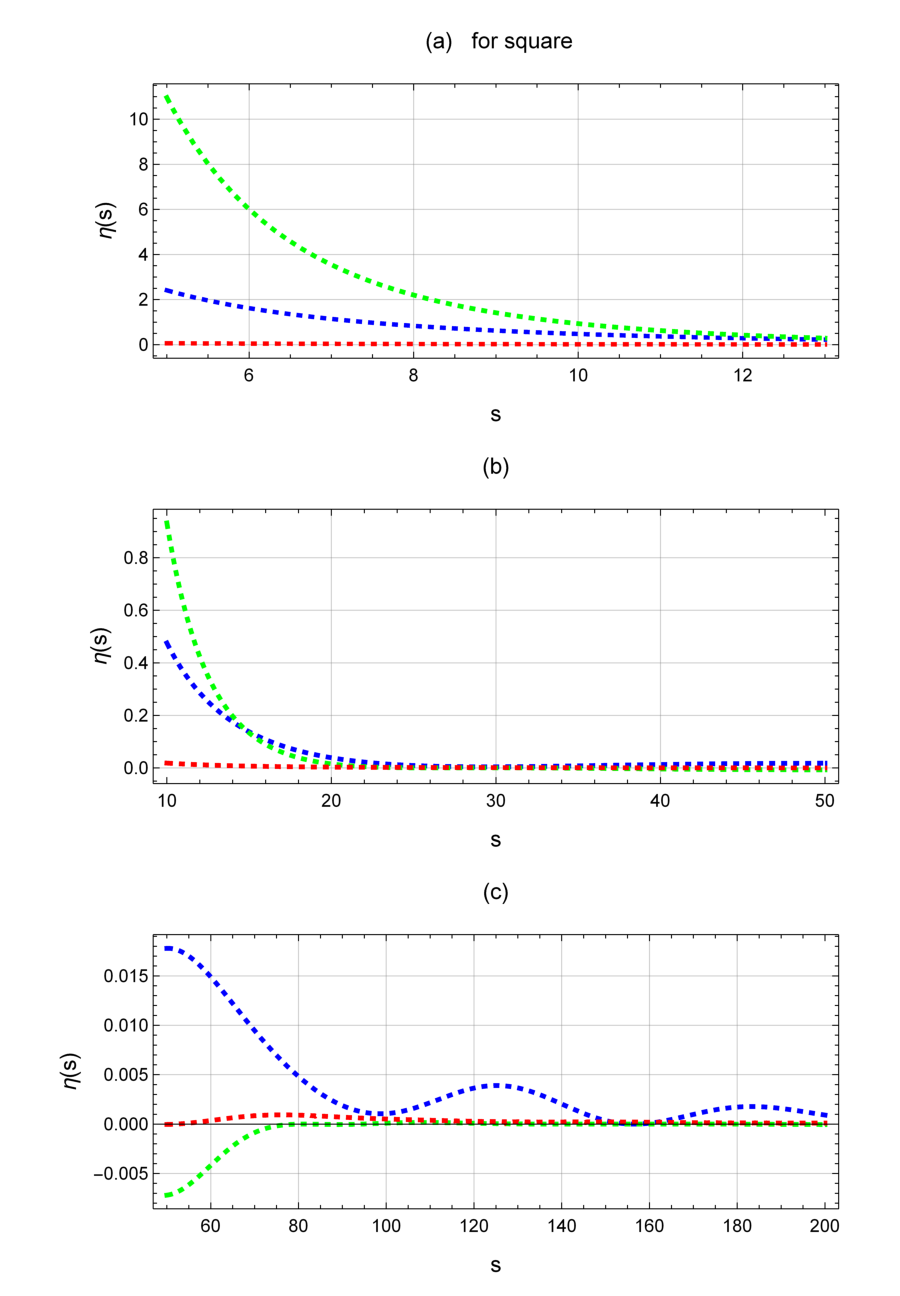}\\
	\caption{  \label{fig9}
	The square configuration.
		(a) $\eta^{FP}$ (green) is dominant at small scales
               and decreases with separation.
		(b) $\eta^{disc}$ (blue) starts to dominate  at middle  scales.
		(c)  $\eta^{disc}$ is oscillatory  at large scales.
              $\eta^I$ (red) is small.
    The behaviors are analogous to those in the conformal line configuration.
	}
\end{figure}

In the square case,
the four galaxies with position $(\mathbf{r, r', r'', r'''})$ form a square.
In the spherical coordinate $(r, \theta, \phi)$,
we put $\mathbf{r''' }=(0,0,0)$, $\mathbf{r''}=(s, 0, 0)$,
$\mathbf{r'}=(s, \pi/2, 0)$,
ie, $\mathbf{r'}$ is on the $x$-axis,
and $\mathbf{r}=(\sqrt{2} s, \pi/4, 0)$.
The configuration is showed in Fig.\ref{fig8}.
Given the square, one has
\bl
\eta^{disc}(s) & = 2 \xi(s)^2 + \xi(\sqrt{2}\, s)^2  ,
 \\
\eta^{FP}(s) & = 8 \xi(s)^2 \xi(\sqrt{2}\, s)
         +4 \xi(s)^3  +4 \xi(\sqrt{2}\, s)^2 \xi(s) ,
\el
\bl
\eta^I(s)&= \frac{k_J^2}{2 \pi A_\mathrm{m}}
 \iiint \xi(\sqrt{x^2+2 s^2-2 s x (\sin \theta \cos \phi + \cos \theta)})
  \xi(\sqrt{x^2+ s^2-2 s x \sin \theta \cos \phi}) \nn \\
& ~~~~~  \times \xi(\sqrt{x^2+ s^2-2 s x \cos \theta})
\xi(x) x^2 \sin \theta \, \mathrm{d} x \mathrm{d} \theta \mathrm{d} \phi.
\el
The resulting $\eta^{disc}$, $\eta^{FP}$ and $\eta^I$
are plotted in Fig.\ref{fig9} for the square configuration.
At small scales $r \lesssim  12$Mpc the Fry-Pebbles part
$\eta^{FP} \propto \xi^3 \propto r^{-3}$ is dominant.
At large scales $r\gtrsim 12$Mpc the disconnected part
$\eta^{disc}$ is dominant and oscillates with a period $\pi/k_0 \sim 60$Mpc,
and the peaks occur at $s \sim 60, 120, 180 h^{-1} \, \mathrm{Mpc}$.
These patterns are analogous to the conformal line case in Fig.\ref{fig7}.
For other parts to show up,  $\eta^{disc}$  can be subtracted off
from the observational data \cite{Philcox2021,Philcox2022}.

\section{Conclusion and Discussion  }

We have presented the equation and solution of the 4PCF of galaxies
in the Gaussian approximation,
as part of a serial analytical study of the $n$PCF of galaxies.
The starting point is the equation \eqref{psifieldequ}
of density fluctuation of a self-gravity fluid  in a static Universe.
The Schwinger functional differentiation technique
that we apply is a powerful tool
in field theory to derive the equation of Green's functions
 when the field equation is given.
The derivation is simpler than working with the Liouville's equation of
 the probability distribution functions in the phase space.
The resulting equation  \eqref{Gaussian4PCF} of 4PCF $\eta$
contains hierarchically 2PCF $\xi$ and 3PCF $\zeta$, as expected.
Given  $\xi$ in \eqref{gsolution} and $\zeta$ in \eqref{GPansatz},
the final equation  \eqref{solve4PCF-1}
of $\eta$ is a closed,  Helmholtz equation with inhomogeneous term
 formed from products  of $\xi$.
Interestingly,  in the Gaussian approximation,
all the partial differential equations of 2PCF, 3PCF and 4PCF
possess a similar structure of the Helmholtz type,
and have two physical parameters: the mass of galaxy and the Jeans wavenumber.
A great advantage of the Gaussian approximation is that
the equation of $n$-point correlation function is linear
in the correlation function itself,
and the analytical solution can be found
to explain the observational data in a transparent manner.

We obtain the  4PCF solution \eqref{4pcfSol}
consisting of four  parts: $\eta= \eta^0_{odd}+ \eta^0_{even} +\eta^{FP}+\eta^I$.
The first two terms $\eta^0_{odd}+ \eta^0_{even}$ form
 the general, homogeneous solution $\eta^0$ \eqref{solv3}.
The most interesting is the parity-odd $\eta^0_{odd}$  \eqref{solv3odd},
which can explain qualitatively the observed parity-odd 4PCF
 from BOSS galaxies \cite{Philcox2021,Philcox2022,Hou2022}.
The parity-even $\eta^0_{even}$ \eqref{solv3even}
contains the disconnected 4PCF $\eta^{disc}$ \eqref{disconn}
that corresponds to  the 4PCF of a Gaussian random process.
$\eta^{FP}$ \eqref{solh} has the same form as the Fry-Peebles ansatz for 4PCF,
$\eta^I$ \eqref{usolution} is the integration of an inhomogeneous term,
and both $\eta^{FP}$ and $\eta^I$  are parity-even.
We also compare the  reduced parity-even 4PCF
with the observation data,
and find that $\eta^{FP}$ is dominant at small scales,
 $\eta^I$ is subdominant,
and  $\eta^0_{odd}+ \eta^0_{even}$
is dominant at large scales,
ie, the signals of parity-odd are more prominent at large scales.

While $\eta^{FP}$ and $\eta^I$ are fully determined by the 2PCF,
 $\eta^0_{odd}$ and $\eta^0_{even}$ are  not.
The amplitudes (via the coefficients $C_{l_1\,  l_2 \, l_3}$)
of  $\eta^0_{odd}$ and $\eta^0_{even}$
will be determined by the boundary condition,
and the radial behaviors of $\eta^0_{odd}$ and $\eta^0_{even}$
are  quasi-periodic oscillatory,
determined by the Jeans wavenumber.
The theory of self-gravity density perturbation itself
does not  predict the boundary condition,
and we take the observational data as the boundary condition.
Statistically, the  configurations of a tetrahedron for 4PCF  in space
and its mirror image (spatial-reflected)
might have an equal probability to exist.
But this is not the case,  as indicated by  the observational data.
The observed  nonvanishing parity-odd 4PCF is presumably
an imprint of parity-violation processes
during some early stages of the expanding Universe,
or generated by some mechanism beyond the self-gravity fluid model in this paper.
This is an issue needing further investigations.

In earlier literature,
in lack of the closed equations of nPCF,
the hierarchical clustering picture was based on
the assumption that the nPCF should be constructed from products
of $(n-1)$ 2PCFs.
Our 4PCF solution shows that the assumption is inadequate to account for
$\eta^I$, $\eta^0_{odd}$ and $\eta^0_{even}$ (except for $\eta^{FP}$),
and particularly misses the parity-odd information on large scales.
Therefore, without the field equation,
purely statistical modelings are not sufficient
to describe the 4PCF  of the system of galaxies.

The Gaussian approximation
is valid at large scales $\gtrsim 1 h^{-1}$ Mpc,
as shown for the 2PCF and 3PCF in our previous work.
For the BOSS CMASS sample \cite{Philcox2022},
the galaxy separations are  $\gg 1 h^{-1}$ Mpc,
the comparison of the 4PCF solution with the observational data shows
the validity of Gaussian approximation.
Improvements can be made beyond the Gaussian approximation
by including higher order terms of fluctuation
in the expansions \eqref{gauexp0} \eqref{kjterm} \eqref{expJ},
and we shall come up with the nonlinear equation of correlation.
It is  expected that the amplitude of 4PCF  will be enhanced at small scales,
while the large-scale behavior will not be substantially altered,
and,  in particular, the parity-odd 4PCF  will remain at large scales.

The Gaussian approximation in our work
 is conceptually different from the Gaussian random process in statistics.
The terminology ``the Gaussian approximation" we use in this paper
is actually adopted from the condensed matter physics,
and is parallel to the Landau-Ginzburg approximation
in the phase transition theory
\cite{Goldenfeld1992,BinneyDowrick1992,Zinn-Justin1996}.
Technically, in our context, the Gaussian approximation is through
the expansions \eqref{gauexp0} \eqref{kjterm} \eqref{expJ}
in terms of fluctuations.
As a working tool,
it represents the next order of approximation
beyond the mean field approximation,
and handles adequately the large-scale fluctuations of the self-gravity fluid.

The  solution \eqref{GPansatz} of 3PCF
in the Gaussian approximation is nonvanishing
for the self-gravity fluid.
Moreover,  the solution \eqref{g4solution} of 4PCF
is beyond a Gaussian random process,
(except for $\eta^{disc}$).
So, both 3PCF and 4PCF demonstrate an important property
that the fluctuations of the self-gravity fluid in the Gaussian approximation
can not be described statistically by a Gaussian random process.
As we understand, this property is due to
the long-range gravity present in the fluid at the fundamental level,
thereby the modes of fluctuations are not independent
and the Gaussian statistics is not appropriate to apply.
Only in the mean field approximation,
the density fluctuations are set to zero,
and consequently all the correlation functions are vanishing.
In this regard, for any consistent treatment of the fluctuations
of the self-gravity fluid, there is no Gaussian statistical limit
in which the two-point  correlation function is nonzero
while the three-point and higher correlation functions vanish.
The above analysis on the structure of the $n$-point correlation functions
shows how the system of galaxies
differs from the fluctuations of the cosmic microwave background radiation (CMB)
\cite{ZhaoZhang2006,XiaZhang2008,XiaZhang2009,Zhang2011,CaiZhang2012}.
In theoretical perspective,
the Maxwell field is a linear, massless field,
so that it has neither coupling between the fluctuation modes \cite{ZhangYePRD2022}
nor Newtonian limit with self-gravity.
Thus, the fluctuations of CMB are statistically independent,
and well described by a Gaussian random process,
so is the relic gravitational wave by the same reason
\cite{AllenRomaro1999,Zhangatel2006,ZhangTongFu2010,WangZhang2019,ZhangWang2018}.

The static model can apply to
the system of the galaxies  distributed in a small redshift range  \cite{Philcox2022},
The evolution effect is small within a small redshift range,
as shown by the evolutionary 2PCF solution in the expanding Universe \cite{ZhangLi2021}.
When observational data are available in future,
one can proceed further to derive the evolutionary equation of 4PCF,
and compare with the data.

Finally, the Newtonian self-gravity fluid as the basic model
is simple enough to work with,
and the equation \eqref{psifieldequ}
contains the density field as the only dynamical variable.
Other possible improvements of the model
are to include the shear tensor and the anisotropic stress in the fluid,
and these will need additional treatments.

{\bf Acknowledgment} : We would thank  Dr. Philcox
     for helpful explanation on the radial bin of the data in his paper.
     Y. Zhang is supported by NSFC Grant No. 11675165, 11961131007, 12261131497,
     and in part by National Key RD Program of China (2021YFC2203100).

\appendix
\numberwithin{equation}{section}

\section{Derivation of the equation of  4PCF in Gaussian approximation}

In this Appendix  we apply the Schwinger functional differentiation
on the ensemble average
of the equation  \eqref{4pcffielda} of the density field,
and derive the equation \eqref{Gaussian4PCF} of 4PCF,
in analogy to the derivation of the equations of 2PCF and 3PCF
\cite{Zhang2007,ZhangMiao2009,ZhangChen2015,
ZhangChenWu2019,ZhangLi2021,WuZhang2022-2,WuZhang2022-8}.

The first term in Eq.(\ref{4pcffielda})
is a functional differentiation with respect to
the external source $J$ three times.
By use of the definition \eqref{def4pcf},  it gives
\be \label{thefirstterm}
\frac{1}{\alpha^3}\frac{\delta^3}{\delta J({\bf r}')
	\delta J({\bf r}'') \delta J({\bf r}''' )}
\nabla^2 \la \psi({\bf r}) \ra {\Big \arrowvert}_{J=0}
=\nabla^2 G^{(4)}( {\bf r} ,  {\bf r}' ,  {\bf r}'', {\bf r}''' ) ,
\ee
where the ordering of $\frac{\delta }{\delta J}$
and   $\nabla^2=\nabla^2_{\mathbf{r}}$ can be exchanged.

The second term  of Eq.(\ref{4pcffielda})
contains $\la -\frac{(\nabla \psi)^2}{\psi} \ra$
which is highly nonlinear due to the factor $\frac{1}{\psi}$.
We write the density field $\psi$ into an averaged density and a  fluctuation
\be \label{expandpsi}
\psi=\la \psi \ra_J + \delta \psi ,
\ee
in the presence of $J$,
and make an expansion as the following
\[
\frac{1}{\psi}=\frac{1}{\langle\psi\rangle_J +\delta\psi}
\simeq\frac{1}{\langle\psi\rangle_J }
\Big(1-\frac{\delta\psi}{\langle\psi_J \rangle}
+ O((\delta\psi)^{2}) \Big) .
\]
Taking its  ensemble average,
using  $\la \delta \psi \ra = 0$,
and dropping the higher $(\delta \psi)^2$ term, we have
\be\label{gauexp0}
\Big\la \frac{1}{\psi} \Big\ra _J
\simeq \frac{1}{\la \psi \ra_J } .
\ee
This is the Gaussian approximation of the density field
that we have adopted in this paper
and in our previous works \cite{Zhang2007,ZhangLi2021}.
Then the second term  of Eq.(\ref{4pcffielda}) is written as
\be\label{gauexp}
\Big\la -\frac{(\nabla \psi)^2}{\psi} \Big\ra_J
 \simeq  -\frac{(\nabla \la \psi \ra_J )^2}{\la \psi \ra_J} .
\ee
In the  following we shall omit the subscript $J$ whenever no confusions arise.
Taking functional differentiation of \eqref{gauexp}
with respect to $J$ once gives
\[
\frac{1}{\alpha } \frac{\delta}{\delta J(\mathbf{r'})}
\Big(-\frac{(\nabla \la \psi \ra)^2}{\la \psi \ra} \Big)
= -\bigg[ -\frac{1}{\la \psi \ra^2} \frac{1}{\alpha}
\frac{\delta \la \psi \ra}{\delta J(\mathbf{r'})} (\nabla \la \psi \ra)^2
+ \frac{2}{\la \psi \ra} \nabla \la \psi \ra \cdot \nabla
\bigg(\frac{1}{\alpha}\frac{\delta \la \psi \ra}{\delta J(\mathbf{r'})} \bigg)
 \bigg] .
\]
Taking the functional differentiation twice gives
\bl \nonumber
& \frac{1}{\alpha^2 } \frac{\delta^2}{\delta J(\mathbf{r'})
    \delta J(\mathbf{r''})} \Big(-\frac{(\nabla \la \psi \ra)^2}{\la \psi \ra}\Big)
\nn \\
=&-\frac{2}{\la \psi \ra^3}\frac{1}{\alpha} \frac{\delta \la \psi \ra}{\delta J(\mathbf{r''})}
 (\nabla \la \psi \ra)^2 \frac{1}{\alpha } \frac{\delta \la \psi \ra}{\delta J(\mathbf{r'})}
+2 \nabla \la \psi \ra \cdot \nabla \bigg(\frac{1}{\alpha} \frac{\delta \la \psi \ra}{\delta J(\mathbf{r''})} \bigg) \frac{1}{\la \psi \ra^2} \frac{1}{\alpha } \frac{\delta \la \psi \ra}{\delta J(\mathbf{r'})} \nonumber \\
&+ \frac{(\nabla \la \psi \ra)^2}{\la \psi \ra^2} \frac{1}{\alpha^2} \frac{\delta^2 \la \psi \ra}{\delta J(\mathbf{r'})\delta J(\mathbf{r''})}
+\frac{2}{\la \psi \ra^2} \frac{1}{\alpha} \frac{\delta \la \psi \ra}{\delta J(\mathbf{r''})}
\nabla \la \psi \ra \cdot \nabla \bigg( \frac{1}{\alpha}
\frac{\delta \la \psi \ra}{\delta J(\mathbf{r'})} \bigg)
\nonumber \\
&-\frac{2}{\la \psi \ra}  \nabla \bigg( \frac{1}{\alpha}
\frac{\delta \la \psi \ra}{\delta J(\mathbf{r''})} \bigg)
\cdot \nabla \bigg( \frac{1}{\alpha}
\frac{\delta \la \psi \ra}{\delta J(\mathbf{r'})} \bigg)
-\frac{2}{\la \psi \ra}  \nabla \la \psi
 \ra \cdot \nabla \bigg( \frac{1}{\alpha^2} \frac{\delta^2
  \la \psi \ra}{\delta J(\mathbf{r'})\delta J(\mathbf{r''})} \bigg) .
\nonumber
\el
Taking the functional differentiation thrice  gives
\ba \label{term2dif3}
&&\frac{1}{\alpha^3 } \frac{\delta^3}{\delta J(\mathbf{r'})
\delta J(\mathbf{r''})\delta J(\mathbf{r'''})}
\Big(-\frac{(\nabla \la \psi \ra)^2}{\la \psi \ra}\Big) \nonumber \\
&=&- \frac{4}{\la \psi \ra^3} \frac{1}{\alpha} \frac{\delta \la \psi \ra}{\delta J(\mathbf{r'''})}
\nabla \la \psi \ra \cdot \nabla \bigg( \frac{1}{\alpha }
\frac{\delta \la \psi \ra}{\delta J(\mathbf{r''})} \bigg)
\frac{1}{\alpha } \frac{\delta \la \psi \ra}{\delta J(\mathbf{r'})}
+\frac{2}{\la \psi \ra^2}  \nabla \bigg( \frac{1}{\alpha} \frac{\delta \la \psi \ra}{\delta J(\mathbf{r'''})} \bigg)  \cdot \nabla \bigg( \frac{1}{\alpha } \frac{\delta \la \psi \ra}{\delta J(\mathbf{r''})} \bigg)
\frac{1}{\alpha } \frac{\delta \la \psi \ra}{\delta J(\mathbf{r'})} \nonumber \\
&&+\frac{2}{\la \psi \ra^2}  \nabla \la \psi \ra \cdot \nabla \bigg(\frac{1}{\alpha^2} \frac{\delta^2 \la \psi \ra}{\delta J(\mathbf{r''}) \delta J(\mathbf{r'''})} \bigg)
\frac{1}{\alpha } \frac{\delta \la \psi \ra}{\delta J(\mathbf{r'})}
+\frac{2}{\la \psi \ra^2}  \nabla \la \psi \ra \cdot \nabla \bigg( \frac{1}{\alpha } \frac{\delta \la \psi \ra}{\delta J(\mathbf{r''})} \bigg)
\frac{1}{\alpha^2} \frac{\delta^2 \la \psi \ra}{\delta J(\mathbf{r'}) \delta J(\mathbf{r'''})} \nonumber \\
&&+\frac{6}{\la \psi \ra^4}  (\nabla \la \psi \ra)^2
\frac{1}{\alpha} \frac{\delta \la \psi \ra}{\delta J(\mathbf{r'''})}
\frac{1}{\alpha } \frac{\delta \la \psi \ra}{\delta J(\mathbf{r''})}
\frac{1}{\alpha } \frac{\delta \la \psi \ra}{\delta J(\mathbf{r'})}
- \frac{4}{\la \psi \ra^3} \nabla \la \psi \ra \cdot \nabla \bigg(
\frac{1}{\alpha} \frac{\delta \la \psi \ra}{\delta J(\mathbf{r'''})} \bigg)
\frac{1}{\alpha } \frac{\delta \la \psi \ra}{\delta J(\mathbf{r''})}
\frac{1}{\alpha } \frac{\delta \la \psi \ra}{\delta J(\mathbf{r'})} \nonumber \\
&&- \frac{2}{\la \psi \ra^3} (\nabla \la \psi \ra)^2
\frac{1}{\alpha^2} \frac{\delta^2 \la \psi \ra}{\delta J(\mathbf{r''}) \delta J(\mathbf{r'''})}
\frac{1}{\alpha } \frac{\delta \la \psi \ra}{\delta J(\mathbf{r'})}
- \frac{2}{\la \psi \ra^3} (\nabla \la \psi \ra)^2 \frac{1}{\alpha } \frac{\delta \la \psi \ra}{\delta J(\mathbf{r''})}
\frac{1}{\alpha^2} \frac{\delta^2 \la \psi \ra}{\delta J(\mathbf{r'}) \delta J(\mathbf{r'''})} \nonumber \\
&&+\frac{1}{\la \psi \ra^4} \bigg[2 \la \psi \ra^2 \nabla \la \psi \ra
\cdot \nabla
\bigg(\frac{1}{\alpha} \frac{\delta \la \psi \ra}{\delta J(\mathbf{r'''})}\bigg)
-2 \la \psi \ra (\nabla \la \psi \ra)^2
\frac{1}{\alpha} \frac{\delta \la \psi \ra}{\delta J(\mathbf{r'''})}\bigg]
\frac{1}{\alpha^2} \frac{\delta^2 \la \psi \ra}{\delta J(\mathbf{r'}) \delta J(\mathbf{r''})} \nonumber \\
&&+\frac{(\nabla \la \psi \ra)^2}{\la \psi \ra^2} \frac{1}{\alpha^3}
\frac{\delta^3 \la \psi \ra}{\delta J(\mathbf{r'}) \delta J(\mathbf{r''}) \delta J(\mathbf{r'''})} \nonumber \\
&&-\frac{4}{\la \psi \ra^3} \frac{1}{\alpha} \frac{\delta \la \psi \ra}{\delta J(\mathbf{r'''})} \frac{1}{\alpha } \frac{\delta \la \psi \ra}{\delta J(\mathbf{r''})}  \nabla \la \psi \ra \cdot \nabla \bigg( \frac{1}{\alpha } \frac{\delta \la \psi \ra}{\delta J(\mathbf{r'})} \bigg)
+ \frac{2}{\la \psi \ra^2} \frac{1}{\alpha^2} \frac{\delta^2 \la \psi \ra}{\delta J(\mathbf{r''}) \delta J(\mathbf{r'''})}  \nabla \la \psi \ra \cdot \nabla \bigg( \frac{1}{\alpha } \frac{\delta \la \psi \ra}{\delta J(\mathbf{r'})} \bigg) \nonumber \\
&&+ \frac{2}{\la \psi \ra^2}\frac{1}{\alpha } \frac{\delta \la \psi \ra}{\delta J(\mathbf{r''})}  \nabla \bigg( \frac{1}{\alpha} \frac{\delta \la \psi \ra}{\delta J(\mathbf{r'''})}\bigg) \cdot \nabla \bigg( \frac{1}{\alpha } \frac{\delta \la \psi \ra}{\delta J(\mathbf{r'})} \bigg)
+ \frac{2}{\la \psi \ra^2}
\frac{1}{\alpha } \frac{\delta \la \psi \ra}{\delta J(\mathbf{r''})}  \nabla \la \psi \ra \cdot
\nabla \bigg(\frac{1}{\alpha^2} \frac{\delta^2 \la \psi \ra}{\delta J(\mathbf{r'}) \delta J(\mathbf{r'''})} \bigg) \nonumber \\
&&+\frac{2}{\la \psi \ra^2} \frac{1}{\alpha} \frac{\delta \la \psi \ra}{\delta J(\mathbf{r'''})}
\nabla \bigg( \frac{1}{\alpha } \frac{\delta \la \psi \ra}{\delta J(\mathbf{r''})} \bigg)
\cdot \nabla \bigg( \frac{1}{\alpha } \frac{\delta \la \psi \ra}{\delta J(\mathbf{r'})} \bigg)
-\frac{2}{\la \psi \ra} \nabla \bigg( \frac{1}{\alpha^2}
\frac{\delta^2 \la \psi \ra}{\delta J(\mathbf{r''}) \delta J(\mathbf{r'''})} \bigg)
\cdot \nabla \bigg( \frac{1}{\alpha } \frac{\delta \la \psi \ra}{\delta J(\mathbf{r'})} \bigg) \nonumber \\
&&-\frac{2}{\la \psi \ra} \nabla \bigg( \frac{1}{\alpha } \frac{\delta \la \psi \ra}{\delta J(\mathbf{r''})} \bigg)
\cdot \nabla \bigg( \frac{1}{\alpha^2} \frac{\delta^2 \la \psi \ra}{\delta J(\mathbf{r'}) \delta J(\mathbf{r'''})}  \bigg) \nonumber \\
&&+\frac{2}{\la \psi \ra^2}\frac{1}{\alpha} \frac{\delta \la \psi \ra}{\delta J(\mathbf{r'''})}   \nabla \la \psi \ra \cdot \nabla \bigg( \frac{1}{\alpha^2} \frac{\delta^2 \la \psi \ra}{\delta J(\mathbf{r'}) \delta J(\mathbf{r''})} \bigg)
-\frac{2}{\la \psi \ra}
\nabla \bigg( \frac{1}{\alpha} \frac{\delta \la \psi \ra}{\delta J(\mathbf{r'''})} \bigg)
\cdot \nabla \bigg( \frac{1}{\alpha^2} \frac{\delta^2 \la \psi \ra}{\delta J(\mathbf{r'}) \delta J(\mathbf{r''})} \bigg) \nonumber \\
&&-\frac{2}{\la \psi \ra}  \nabla \la \psi \ra \cdot \nabla \bigg( \frac{1}{\alpha^3}
\frac{\delta^3 \la \psi \ra}{\delta J(\mathbf{r'})
\delta J(\mathbf{r''})\delta J(\mathbf{r'''})} \bigg).
\ea
Setting  $J=0$ in \eqref{term2dif3},
we get the contribution  of the second term of Eq.\eqref{4pcffielda}
as the following
\bl \label{thesecondterm}
& \frac{1}{\alpha^3} \frac{\delta^3}{\delta J(\mathbf{r'})
\delta J(\mathbf{r''}) \delta J(\mathbf{r'''})}
\Big\la -\frac{(\nabla \psi)^2}{\psi} \Big\ra
  {\Big \arrowvert}_{J=0}
  \nonumber \\
= & \bigg[ \frac{2}{\la \psi \ra^2}  \nabla \bigg(
\frac{1}{\alpha} \frac{\delta \la \psi \ra}{\delta J(\mathbf{r'''})} \bigg)
\cdot \nabla \bigg( \frac{1}{\alpha } \frac{\delta \la \psi \ra}{\delta J(\mathbf{r''})} \bigg)
\frac{1}{\alpha } \frac{\delta \la \psi \ra}{\delta J(\mathbf{r'})}
+ \frac{2}{\la \psi \ra^2}\frac{1}{\alpha } \frac{\delta \la \psi \ra}{\delta J(\mathbf{r''})}
\nabla \bigg( \frac{1}{\alpha} \frac{\delta \la \psi \ra}{\delta J(\mathbf{r'''})}\bigg)
\cdot \nabla \bigg( \frac{1}{\alpha } \frac{\delta \la \psi \ra}{\delta J(\mathbf{r'})} \bigg)
\nonumber \\
&+\frac{2}{\la \psi \ra^2} \frac{1}{\alpha} \frac{\delta \la \psi \ra}{\delta J(\mathbf{r'''})}
\nabla \bigg( \frac{1}{\alpha } \frac{\delta \la \psi \ra}{\delta J(\mathbf{r''})} \bigg)
\cdot \nabla \bigg( \frac{1}{\alpha } \frac{\delta \la \psi \ra}{\delta J(\mathbf{r'})} \bigg)
-\frac{2}{\la \psi \ra} \nabla \bigg(
\frac{1}{\alpha^2} \frac{\delta^2 \la \psi \ra}{\delta J(\mathbf{r''}) \delta J(\mathbf{r'''})} \bigg)
\cdot \nabla \bigg( \frac{1}{\alpha } \frac{\delta \la \psi \ra}{\delta J(\mathbf{r'})} \bigg) \nonumber \\
&-\frac{2}{\la \psi \ra} \nabla \bigg(
\frac{1}{\alpha } \frac{\delta \la \psi \ra}{\delta J(\mathbf{r''})} \bigg)
\cdot \nabla \bigg( \frac{1}{\alpha^2}
\frac{\delta^2 \la \psi \ra}{\delta J(\mathbf{r'}) \delta J(\mathbf{r'''})}  \bigg)
-\frac{2}{\la \psi \ra}
\nabla \bigg( \frac{1}{\alpha} \frac{\delta \la \psi \ra}{\delta J(\mathbf{r'''})} \bigg)
\cdot \nabla \bigg( \frac{1}{\alpha^2}
\frac{\delta^2 \la \psi \ra}{\delta J(\mathbf{r'})
 \delta J(\mathbf{r''})} \bigg) \bigg] {\Bigg \arrowvert}_{J=0}
 \nonumber \\
= & 2 G^{(2)}(\mathbf{r, r'}) \nabla G^{(2)}(\mathbf{r, r''})
\cdot \nabla G^{(2)}(\mathbf{r, r'''})
+ 2 G^{(2)}(\mathbf{r, r''})
\nabla G^{(2)}(\mathbf{r, r'''})
\cdot \nabla G^{(2)}(\mathbf{r, r'})
\nonumber \\
&+2 G^{(2)}(\mathbf{r, r'''})
\nabla G^{(2)}(\mathbf{r, r'})
\cdot \nabla G^{(2)}(\mathbf{r, r''})
-2 \nabla G^{(2)}(\mathbf{r, r'}) \cdot
\nabla G^{(3)}(\mathbf{r, r'', r'''})  \nonumber \\
&-2 \nabla G^{(2)}(\mathbf{r, r''})
\cdot \nabla G^{(3)}(\mathbf{r, r', r'''})
-2
\nabla G^{(2)}(\mathbf{r, r'''})
\cdot \nabla G^{(3)}(\mathbf{r, r', r''}) ,
\el
where we have used $\la \psi \ra_{J=0} =\psi_0 = 1 $,
$\nabla \la \psi \ra_{J=0} = 0$,  Eq.\eqref{G2cal}, and  Eq.\eqref{G3cal}.

The third term of Eq.(\ref{4pcffielda}) is due to
the Jeans term $k_J^2 \la \psi^2 \ra$.
Expanding $\psi^2$  gives
\be \label{kjterm}
k_J^2 \la  \psi^2 \ra=k_J^2  \la \big(\la \psi \ra + \delta \psi \big)^2 \ra
= k_J^2 \la \psi \ra^2+k_J^2 \la \delta \psi \delta \psi \ra
\simeq k_J^2 \la \psi \ra^2,
\ee
where $\la \delta \psi \ra \equiv 0$ has been used,
and $(\delta \psi)^2$ has been dropped in the Gaussian approximation.
Taking the functional differentiation three times gives
\bl \label{term3dif3}
& \frac{1}{\alpha^3 } \frac{\delta^3 (k_J^2 \la \psi \ra^2)}
  {\delta J(\mathbf{r'}) \delta J(\mathbf{r''}) \delta J(\mathbf{r'''})}
    \nonumber \\
& =  2 k_J^2  \Big(  \frac{1}{\alpha^2}
\frac{\delta^2 \la \psi \ra}{\delta J(\mathbf{r'})\delta J(\mathbf{r'''})}
\frac{1}{\alpha } \frac{\delta \la \psi \ra}{\delta J(\mathbf{r''})}
+\frac{1}{\alpha } \frac{\delta \la \psi \ra}{\delta J(\mathbf{r'})}
 \frac{1}{\alpha^2} \frac{\delta^2 \la \psi \ra}{\delta J(\mathbf{r''})
 \delta J(\mathbf{r'''})} \nonumber \\
& + \frac{1}{\alpha} \frac{\delta \la \psi \ra}{\delta J(\mathbf{r'''})}
\frac{1}{\alpha^2 } \frac{\delta^2 \la \psi \ra}{\delta J(\mathbf{r'})
\delta J(\mathbf{r''})}
+ \la \psi \ra \frac{1}{\alpha^3 }
\frac{\delta^3 \la \psi \ra}{\delta J(\mathbf{r'})\delta J(\mathbf{r''})
\delta J(\mathbf{r'''}) } \Big).
\el
Setting  $J=0$ in the above gives the contribution of  the third term
\bl \label{thethirdterm}
 \frac{1}{\alpha^3} \frac{\delta^3 ( k_J^2 \la \psi^2 \ra ) }{\delta J(\mathbf{r'})
 \delta J(\mathbf{r''}) \delta J(\mathbf{r'''})}
 {\Big \arrowvert}_{J=0}
= & 2 k_J^2 \Big(G^{(3)}(\mathbf{r, r', r'''}) G^{(2)}(\mathbf{r, r''})
+G^{(3)}(\mathbf{r, r'', r'''}) G^{(2)}(\mathbf{r, r'})
\nn \\
& +G^{(3)}(\mathbf{r, r', r''}) G^{(2)}(\mathbf{r, r'''})
 +   G^{(4)}(\mathbf{r, r', r'', r'''})\Big).
\el

The fourth term of Eq.(\ref{4pcffielda})
is the external source term $J \la  \psi^2 \ra_J$,
which, in the Gaussian approximation, is written  as
\bl \label{expJ}
 J  \la \psi^2 \ra_J \simeq  J \la \psi \ra_J ^2 .
\el
Taking the functional differentiation three times
and using $\frac{\delta J(\mathbf{r})}{\delta J(\mathbf{r'})}
= \delta^{(3)}(\mathbf{r-r'})$,
we get
\bl  \label{term4dif3}
&\frac{1}{\alpha^3 } \frac{\delta^3 (J \la \psi \ra^2)}{\delta J(\mathbf{r'})
 \delta J(\mathbf{r''})\delta J(\mathbf{r'''})}
=  2 \bigg( \frac{1}{\alpha} \frac{\delta \la \psi \ra}{\delta J(\mathbf{r'''})}
\frac{1}{\alpha } \delta^{(3)}(\mathbf{r-r'})
\frac{1}{\alpha} \frac{\delta \la \psi \ra}{\delta J(\mathbf{r''})}
+\la \psi \ra \frac{1}{\alpha } \delta^{(3)}(\mathbf{r-r'})
\frac{1}{\alpha^2} \frac{\delta^2 \la \psi \ra}{\delta J(\mathbf{r''})
 \delta J(\mathbf{r'''})} \nonumber\\
&~~~ + \frac{1}{\alpha} \delta^{(3)}(\mathbf{r-r''})
\frac{1}{\alpha} \frac{\delta \la \psi \ra}{\delta J(\mathbf{r'''})}
\frac{1}{\alpha } \frac{\delta \la \psi \ra}{\delta J(\mathbf{r'})}
+ \frac{1}{\alpha} \delta^{(3)}(\mathbf{r-r''}) \la \psi \ra
\frac{1}{\alpha^2} \frac{\delta^2 \la \psi \ra}{\delta J(\mathbf{r'})\delta J(\mathbf{r'''})} \nonumber \\
&~~~ + \frac{1}{\alpha} \delta^{(3)} (\mathbf{r-r'''})
\frac{1}{\alpha} \frac{\delta \la \psi \ra}{\delta J(\mathbf{r''})}
\frac{1}{\alpha } \frac{\delta \la \psi \ra}{\delta J(\mathbf{r'})}
+ J \frac{1}{\alpha^2} \frac{\delta^2 \la \psi \ra}{\delta J(\mathbf{r''}) \delta J(\mathbf{r'''})}
\frac{1}{\alpha } \frac{\delta \la \psi \ra}{\delta J(\mathbf{r'})}
+ J \frac{1}{\alpha} \frac{\delta \la \psi \ra}{\delta J(\mathbf{r''})}
\frac{1}{\alpha^2} \frac{\delta^2 \la \psi \ra}{\delta J(\mathbf{r'}) \delta J(\mathbf{r'''})} \nonumber \\
&~~~ + \frac{1}{\alpha} \delta^{(3)} (\mathbf{r-r'''})  \la \psi \ra
\frac{1}{\alpha^2} \frac{\delta^2 \la \psi \ra}{\delta J(\mathbf{r'})\delta J(\mathbf{r''})}
+ J \frac{1}{\alpha} \frac{\delta \la \psi \ra}{\delta J(\mathbf{r'''})}  \frac{1}{\alpha^2} \frac{\delta^2 \la \psi \ra}{\delta J(\mathbf{r'})\delta J(\mathbf{r''})}
+ J \la \psi \ra \frac{1}{\alpha^3} \frac{\delta^3 \la \psi \ra}{\delta J(\mathbf{r'})\delta J(\mathbf{r''}) \delta J(\mathbf{r'''})} \bigg). \nonumber \\
\el
Taking $J=0$ in the above gives the contribution of the fourth term
as the following
\bl \label{theforthterm}
 \frac{1}{\alpha^3} \frac{\delta^3 \la J \psi^2 \ra    }{\delta J(\mathbf{r'})
 \delta J(\mathbf{r''}) \delta J(\mathbf{r'''})}
{\Big |}_{J=0}
& =  \frac{2}{\alpha} \delta^{(3)}(\mathbf{r-r'})
  \Big( G^{(2)}(\mathbf{r, r''}) G^{(2)}(\mathbf{r, r'''})
+   G^{(3)}(\mathbf{r, r'', r'''}) \Big) \nonumber\\
& + \frac{2}{\alpha} \delta^{(3)}(\mathbf{r-r''})
   \Big( G^{(2)}(\mathbf{r, r'}) G^{(2)}(\mathbf{r, r'''})
+   G^{(3)}(\mathbf{r, r', r'''}) \Big) \nonumber \\
& + \frac{2}{\alpha} \delta^{(3)} (\mathbf{r-r'''})
\Big( G^{(2)}(\mathbf{r, r'}) G^{(2)}(\mathbf{r, r''})
+    G^{(3)}(\mathbf{r, r', r''}) \Big) .
\el
Putting the four terms
  \eqref{thefirstterm} \eqref{thesecondterm}
\eqref{thethirdterm} \eqref{theforthterm}
into Eq.\eqref{4pcffielda},
we arrive at Eq.\eqref{Gaussian4PCF} of $G^{(4)}$.

\end{document}